\documentclass[10pt,journal,compsoc]{IEEEtran}

\ifCLASSINFOpdf
\else
\fi

\usepackage{cite}
\usepackage{amsmath,amssymb,amsfonts}
\usepackage{algorithmic}
\usepackage[linesnumbered,algoruled,boxed,lined]{algorithm2e}
\usepackage{graphicx}
\usepackage{textcomp}
\usepackage{xcolor}
\usepackage{multirow}
\usepackage{bm}
\usepackage{booktabs}
\usepackage{ntheorem}
\usepackage{tabularx}
\usepackage{subfigure}
\usepackage{makecell}
\usepackage{bbding}
\usepackage{mathrsfs}
\usepackage{ulem}

\theoremseparator{.}
\newskip\theorempreskipamount
\newskip\theorempostskipamount

\theorembodyfont{}

\allowdisplaybreaks[4]

\begin{document}

\title{RIS-assisted Data Collection and Wireless Power Transfer in Low-altitude Wireless Networks}

\author{Wenwen Xie,
        Geng~Sun,~\IEEEmembership{Senior Member,~IEEE,}
        Jiahui~Li,
        Jiacheng Wang,
       Yinqiu Liu,\\
        Dusit Niyato,~\IEEEmembership{Fellow,~IEEE,}
        and Dong In Kim,~\IEEEmembership{Fellow,~IEEE,}
        Shiwen Mao,~\IEEEmembership{Fellow,~IEEE}
        \thanks
        {
        \par Wenwen~Xie and Jiahui Li are with the College of Computer Science and Technology, Jilin University, Changchun 130012, China~(e-mails: xieww22@mails.jlu.edu.cn, lijiahui@jlu.edu.cn).
        \par Geng~Sun is with the College of Computer Science and Technology, Jilin University, Changchun 130012, China, and also with the College of Computing and Data Science, Nanyang Technological University, Singapore 639798 (e-mail: sungeng@jlu.edu.cn).
        \par Jiacheng Wang, Yinqiu Liu, and Dusit Niyato are with the College of Computing and Data Science, Nanyang Technological University, Singapore 639798 (e-mails: jiacheng.wang@ntu.edu.sg, yinqiu001@ntu.edu.sg, dniyato@ntu.edu.sg).
        \par Dong In Kim is with the Department of Electrical and Computer Engineering, Sungkyunkwan University, Suwon 16419, South Korea (e-mail:dongin@skku.edu).
        \par Shiwen Mao with the Department of Electrical and Computer Engineering, Auburn University, Auburn 36830, USA (e-mail: smao@ieee.org).
       \par (\textit{Corresponding authors: Geng Sun and Jiahui Li.})
        }

        \thanks{Part of this paper appeared in IEEE GLOBECOM 2024~\cite{Xie2024a}.}
        
}

\IEEEtitleabstractindextext{%
\begin{abstract}
Due to their flexible deployment capability, low-altitude wireless networks (LAWNs) have become effective solutions for collecting data from low-power Internet-of-Things devices (IoTDs) in remote areas with limited communication infrastructure. However, some outdoor IoTDs deployed in such areas face both energy constraints and low-channel quality challenges, making it challenging to ensure timely data collection from these IoTDs in LAWNs. In this work, we investigate a reconfigurable intelligent surface (RIS)-assisted uncrewed aerial vehicle (UAV)-enabled data collection and wireless power transfer system in LAWN. Specifically, IoTDs first harvest energy from a low-altitude UAV, and then upload their data to the UAV by applying the time division multiple access (TDMA) protocol, supported by an RIS to improve the channel quality. To maintain satisfactory data freshness of the IoTDs and save energy for an energy-constrained UAV, we aim to minimize the age of information (AoI) and energy consumption of the UAV by jointly optimizing the RIS phase shits, UAV trajectory, charging time allocation, and binary IoTD scheduling. Given that the optimization problem is a mixed-integer non-convex problem with dynamic characteristics, we propose a deep reinforcement learning (DRL)-based approach, namely the alternating optimization-improved parameterized deep Q-network (AO-IPDQN). Specifically, considering that RIS typically contains a large number of reflecting elements, we first adopt an alternating optimization (AO) method to optimize the RIS phase shifts to reduce the dimension of the action space. Then, we propose the improved parameterized deep Q-network (IPDQN) method to deal with the hybrid action space involving the continuous UAV trajectory, charging time allocation, and binary IoTD scheduling, which utilizes the prioritized experience replay (PER) mechanism and genetic algorithm (GA) to improve the learning efficiency and the exploration capability of the original DRL method. Simulation results indicate that AO-IPDQN approach achieves excellent performance relative to multiple comparison methods across various simulation scenarios.



\end{abstract}

}

\maketitle

\section{Introduction}
\label{Introduction}

\par With the growing maturity of mobile communications and airspace management, low-altitude wireless networks (LAWNs) based on uncrewed aerial vehicles (UAVs) and airships have attracted significant attention~\cite{11017717}. In particular, due to their cost-effectiveness and rapid deployment, LAWNs show great potential for providing communication services in infrastructure-limited remote areas. Specifically, with the proliferation of the Internet of Things (IoT), numerous low-power IoT devices (IoTDs) have been deployed in rural or mountainous regions for environmental monitoring~\cite{Shahab2024}, and they should upload the monitored data to decision centers for analysis. However, conventional terrestrial networks typically struggle to support long-distance and frequent data transmission due to high deployment costs, lack of infrastructure, and limited IoTD transmit power. In this case, LAWNs are regarded promising solutions to address these limitations~\cite{Wei2025}. Specifically, as key low-altitude platforms in LAWNs, UAVs with high maneuverability can approach IoTDs to perform data collection tasks and then process them locally or relay data to a remote decision center, thus reducing the transmit distance of IoTD data and improving collection efficiency~\cite{10922390,Lu2025}. Notably, ensuring the timeliness of collected IoTD data is essential, as excessive latency or outdated data can reduce decision accuracy and cause economic losses. Therefore, optimizing the age of information (AoI) in UAV-enabled data collection systems in LAWNs is crucial~\cite{Kaul2012,Yang2025}.

\par Given that the UAV communication is heavily dependent on the line-of-sight (LoS) links, improving the channel quality is essential for boosting the AoI performance of LAWNs, especially in the presence of possible obstacles~\cite{Pan2023,Meng2022}. In this case, a reconfigurable intelligent surface (RIS) can be introduced to improve channel conditions by reshaping the wireless channel~\cite{10858311}. In particular, an RIS is composed of numerous passive reflecting elements, where each element can manipulate the phase and amplitude to passively redirect the incoming signal~\cite{Shao2024,Ahmed2025}. As such, the signal propagation path is modified by the RIS to establish the virtual LoS link, thereby mitigating the negative influence of passible obstacles in LAWNs. Moreover, RIS is more energy-saving than traditional relay-assisted method due to its passive beamforming nature~\cite{Wu2023}. TThus, deploying RIS enables the UAV to establish higher-quality links with ground IoTDs while improving AoI performance in a more energy-efficient manner.

\par Moreover, the limited energy of IoTDs poses another significant challenge to the AoI performance of data collection systems in LAWNs. This is because most IoTDs are typically battery-powered in remote areas, such that continuous data transmission and sensing operations can rapidly deplete their energy, potentially leading to device shutdowns due to energy exhaustion. In this case, IoTDs should be provided with sustainable and sufficient energy support to ensure successful data transmission for better AoI performance. Currently, several charging methods have been adopted to charge the IoTDs, such as renewable energy charging ($\textit{e.g.}$, solar energy and wind energy)~\cite{Du2023}. However, due to the time-varying and uncertainty of renewable energy sources, the deployed IoTDs face challenges in obtaining a reliable energy supply in this case~\cite{Li2019a}. Thus, it is imperative to find a new way to charge IoTDs efficiently and reliably.

\par Employing radio frequency (RF)-based wireless power transfer offers a potential solution to mitigate the energy constraints encountered by battery-driven IoTDs in LAWNs~\cite{Xie2025}. Specifically, UAVs equipped with RF transmitters can simultaneously perform data collection tasks and provide wireless charging for spatially distributed IoTDs~\cite{Li2025,Li2023,Shi2024}. This approach offers three key advantages. \textit{First}, thanks to the cost-effectiveness of the UAV, using it for wireless charging is more economical compared to constructing fixed charging infrastructures for outdoor IoTDs~\cite{Liu2021}. \textit{Second}, a high-altitude UAV can provide broader wireless coverage, which can effectively expand the power transfer range for ground IoTDs compared to terrestrial charging infrastructures. \textit{Finally}, with its controllability and maneuverability, the UAV is able to dynamically adapt its flight path according to the system state, which enhances the wireless link quality and boosts the energy harvesting efficiency of IoTDs~\cite{Liang2025}. Motivated by this, we aim to design a UAV-enabled data collection and wireless power transfer system with the assistance of an RIS in the LAWN.

\par Optimizing the performance of such a system faces several significant challenges. \textit{First}, to improve the AoI performance, the UAV should quickly fly to the appropriate area. However, as an energy-constrained platform, the excessive flight speed of the UAV leads to higher energy consumption, which impacts the system sustainability~\cite{Zhang2024}. \textit{Second}, the UAV mobility causes drastic fluctuations in channel conditions, while the IoTD data uploads are influenced by various factors, which results in the high dynamic and uncertainty characteristics within the system. \textit{Third}, the system operates over a long period, where each decision has a profound impact on subsequent system performance, indicating that the system should considers both the short-term and long-term benefits. \textit{Finally}, improving the system performance requires the joint optimization of multiple discrete or continuous variables, such as the RIS phase shifts and UAV trajectory, which are highly coupled and further increase the complexity of the system optimization. Therefore, it is crucial to propose an approach capable of efficiently optimizing mixed variables in such a dynamic system.

\par To address this, we propose a novel online deep reinforcement learning (DRL)-based approach to minimize both AoI and UAV energy consumption in an RIS-assisted UAV-enabled data collection and wireless power transfer system in the LAWN. Our main contributions are as follows:

\begin{itemize}

    \item \textit{RIS-assisted UAV-enabled Data Collection and Wireless Power Transfer System:} We consider employing a UAV to simultaneously perform data collection and power transfer with the assistance of an RIS in LAWN for achieving better information freshness. Specifically, the UAV adopts RF-based power transfer method to charge multiple energy-limited low-power IoTDs with RIS support, following which these IoTDs upload their data to the UAV by using the harvested energy via the time-division multiple access (TDMA) protocol. This system is suitable for practical applications in remote areas, such as environmental monitoring.
    
    \item \textit{Formulation of Dynamic Multi-objective Optimization Problem:} Considering the importance of information freshness for the decision-making process in LAWN applications and the limited energy of the UAV, we formulate an multi-objective optimization problem by jointly optimizing the RIS phase shifts, UAV trajectory, charging time allocation and IoTD scheduling to minimize the AoI of IoTDs and the energy consumption of the UAV. Given that the optimization problem poses the mixed-integer and dynamic characteristics, solving it is challenging.

    \item \textit{Improved DRL-based Approach:} Given the advantages of DRL algorithms in handling dynamic optimization problems, we propose a DRL-based approach for tackling the formulated optimization problem, namely the alternating optimization-improved parameterized deep Q-network (AO-IPDQN) approach. Specifically, we first use the alternating optimization (AO) method to optimize the RIS phase shifts, thereby reducing the action space by $M_{r} \times M_{c}$ dimensions, where $M_{r}$ and $M_{c}$ correspond to the number of RIS elements in each row and column, respectively. Then, we propose an improved parameterized deep Q-network (IPDQN) method, which is designed to effectively handle hybrid action spaces, to optimize the UAV trajectory, charging time, and IoTD scheduling. This method combines the prioritized experience replay (PER) mechanism and the evolutionary mechanism of the genetic algorithm (GA), which can further improve the learning efficiency and exploration capabilities of the method.
    
    \item \textit{Simulations and Analyses:} Simulation results demonstrate that the proposed AO-IPDQN approach achieves beneficial AoI performance and energy efficiency, and outperforms various benchmarks. In particular, the proposed AO-IPDQN approach maintains robust performance even in challenging scenarios, such as when high minimum data size threshold is required or the energy buffer capacity of IoTDs is limited. Moreover, UAV flight trajectories under different minimum data size thresholds indicate that the AO-IPDQN approach can dynamically design more reasonable flight paths according to system requirements and environment conditions.
    
\end{itemize}

\par The rest of this work is organized as follows. Section~\ref{sec:Related Work} introduces the related works. Section~\ref{sec:Models and Preliminaries} provides the system model. Section~\ref{sec:Problem Formulation and Analysis} shows the problem formulation and analysis. Section~\ref{sec:Proposed Soution} proposes the AO-IPDQN approach to solve the formulated problem. Simulations are conducted in Section~\ref{sec:Simulation Results And Analysis} to confirm the effectiveness of the proposed approach. Finally, a summary is given in Section~\ref{sec:Conclusion}.

\section{Related Work}
\label{sec:Related Work}
\par In this section, we present related works on data collection in LAWNs to illustrate the novelty of our work.

\subsection{Data Collection in LAWNs}

\par Due to the high cost-effectiveness and scalable data collection range of UAVs, they are commonly used to collect data from IoTDs in LAWNs. For example, in~\cite{Zhu2023}, a UAV performed data collection from cluster-based IoTDs, after which the collected data were delivered to the base station for subsequent analysis. In~\cite{Xiao2024}, the authors further investigated data collection using multiple UAVs to enhance acquisition efficiency and addressed the coordination issues among the UAVs. Moreover, some existing works focused on the limited energy of IoTDs, allowing UAVs to utilize wireless charging methods to provide timely energy replenishment for IoTDs while collecting data. For example, in~\cite{Liang2025}, a UAV-assisted secure data collection and power transfer scheme was explored to counter malicious interference in the IoT environment and tackle the energy-constraint challenge of low-power IoTDs. In~\cite{Messaoudi2024}, the authors addressed the risk of UAV energy depletion by deploying an unmanned ground vehicle to recharge the UAV when the UAV performs data collection and power transfer tasks, which significantly extends the operation time of the system.

\par Some existing works explored the effectiveness of UAV-enabled data collection systems in LAWNs with RIS support due to the advantages of RIS in improving the channel quality. For example, in~\cite{Fan2023}, attention was given to the challenges posed by dense urban building on the UAV data collection, thereby deploying an RIS at a high altitude to improve the wireless propagation environment. The authors demonstrated that deploying an RIS can significantly improve the data freshness of IoTDs compared to the situations without an RIS. Moreover, an RIS-assisted scheme for efficient and secure data collection from IoTDs by UAVs was investigated under conditions where direct transmission links are blocked and a jammer is present~\cite{Wang2025a}. In addition, in~\cite{Jiang2023a}, the authors utilized a UAV to mount an RIS, forming an aerial RIS to collect data from remote IoTDs, which provided greater flexibility for IoT systems.


\par However, some of the aforementioned works did not consider the energy limitations of IoTDs, which could affect their effectiveness in remote areas where wired charging is not feasible. Moreover, in rural or mountainous regions, dense obstacles such as houses and trees can reduce the quality of air-to-ground channels in LAWNs, thereby negatively impacting data collection efficiency. As such, investigating an RIS-assisted UAV-enabled data collection and wireless power transfer system is practical for remote areas.

\subsection{Performance Metric for Data Collection in LAWNs}

\par Different optimization objectives reflect different system priorities. Currently, research on data collection using UAVs in LAWNs tends to focus on metrics related to data transmission, such as AoI and transmission time. For example, In~\cite{Gao2023}, the authors focused on reducing both the peak and average AoI by coordinating the UAV flight path with the sequence in which IoTDs are scheduled. Moreover, the data collection requirements in disaster scenarios were addressed by minimizing the weighted AoI through jointly designing the UAV flight path, data collection time, and charging time, ensuring that both high-priority and low-priority sensor nodes have data uploaded in time~\cite{Fu2025}. In addition, the authors in~\cite{Gao2023a} investigated urgent data collection tasks in widely distributed IoTD scenarios and developed methods that coordinate task allocation, UAV trajectory, and speed to achieve lower maximum task completion time. 

\par Furthermore, some existing works highlighted the importance of saving UAV energy in UAV-enabled data collection systems in LAWNs. For example, the authors in~\cite{Liu2025} addressed minimization of average AoI by coordinating the transmit power of sensor nodes on islands, the structure of island clusters, and the UAV flight path while considering UAV energy limitations. In~\cite{Deng2024}, the authors designed an aerial relay-based data transmission system and minimized AoI by optimizing UAV hovering locations, hovering durations, and pairing strategies under the constraints of UAV energy consumption. In addition, the authors in~\cite{Huo2024} focused on reducing UAV energy consumption through joint design of task completion time, UAV path, and IoTD scheduling in the UAV-assisted data collection system.

\par However, most of the aforementioned works either focus solely on optimizing data transmission performance or treat UAV energy consumption as a constraint while improving transmission performance. Since the UAV operates under energy limitations, efforts to lower UAV energy consumption while maintaining fresh data offer a pathway to jointly improve the transmission performance and energy efficiency of LAWNs.

\subsection{Optimization Algorithms for Data Collection in LAWNs}

\par Currently, various optimization algorithms are proposed in UAV-enabled data collection systems in LAWNs, such as convex optimization or evolutionary algorithms. For example, the authors in~\cite{Liu2024} deployed multiple charging stations to supply energy to the UAV and enhance the sustainability of the data uploading system, and proposed a convex optimization combined with a greedy approach to optimize the UAV flight path and charging station placement to reduce the AoI metric. Moreover, the authors in~\cite{Liu2025a} explored the feasibility of UAV-assisted agricultural data collection scenarios and proposed an enhanced multi-objective artificial hummingbird algorithm, which jointly design of the UAV flight path and transmit power of IoTDs for minimizing the total energy consumption of the system. Similarly, in~\cite{AbdelBasset2024}, a differential evolution algorithm and a gradient-based optimizer was proposed for UAV-enabled IoT data collection systems to achieve the energy consumption minimization.

\par DRL algorithms are increasingly utilized to UAV-enabled data collection systems in LAWNs due to the powerful adaptability of DRL. For example, in~\cite{10625275}, a multi-agent deep Q-network (MADQN) algorithm was employed for minimizing AoI in a multi-UAV-enabled data collection and wireless power transfer system. In~\cite{Fan2023}, the authors adopted an improved soft actor-critic (SAC) algorithm to jointly design the UAV flight path, IoTD scheduling, and RIS phase shifts in the urban scenario to minimize AoI. Furthermore, the authors in~\cite{Yi2023} introduced a multi-task transfer DRL method that enables policies to be reused in environments sharing similar characteristics, thus eliminating the need for retraining, and optimized the UAV flight path, scheduling of transmissions, and energy recharging to reduce both the average total AoI and charging cost.

\par However, conventional optimization algorithms typically rely on precise prior knowledge, which is challenging to obtain in dynamic environments. Moreover, most of the aforementioned DRL-based algorithms only focus on discrete action spaces. For optimization problems involving mixed-variable, the DRL algorithm above approximates discrete variables as continuous ones~\cite{Fan2023}, which may affect the accuracy of the optimization solution.

%
%
\section{System Model}
\label{sec:Models and Preliminaries}
\par In this section, we first introduce the components and operating process of the considered system. Subsequently, we model the channel and transmission process between the UAV and IoTDs. Finally, we present the energy consumption model and AoI model.

\begin{figure}[t]
\centering
\includegraphics[width=3.5in]{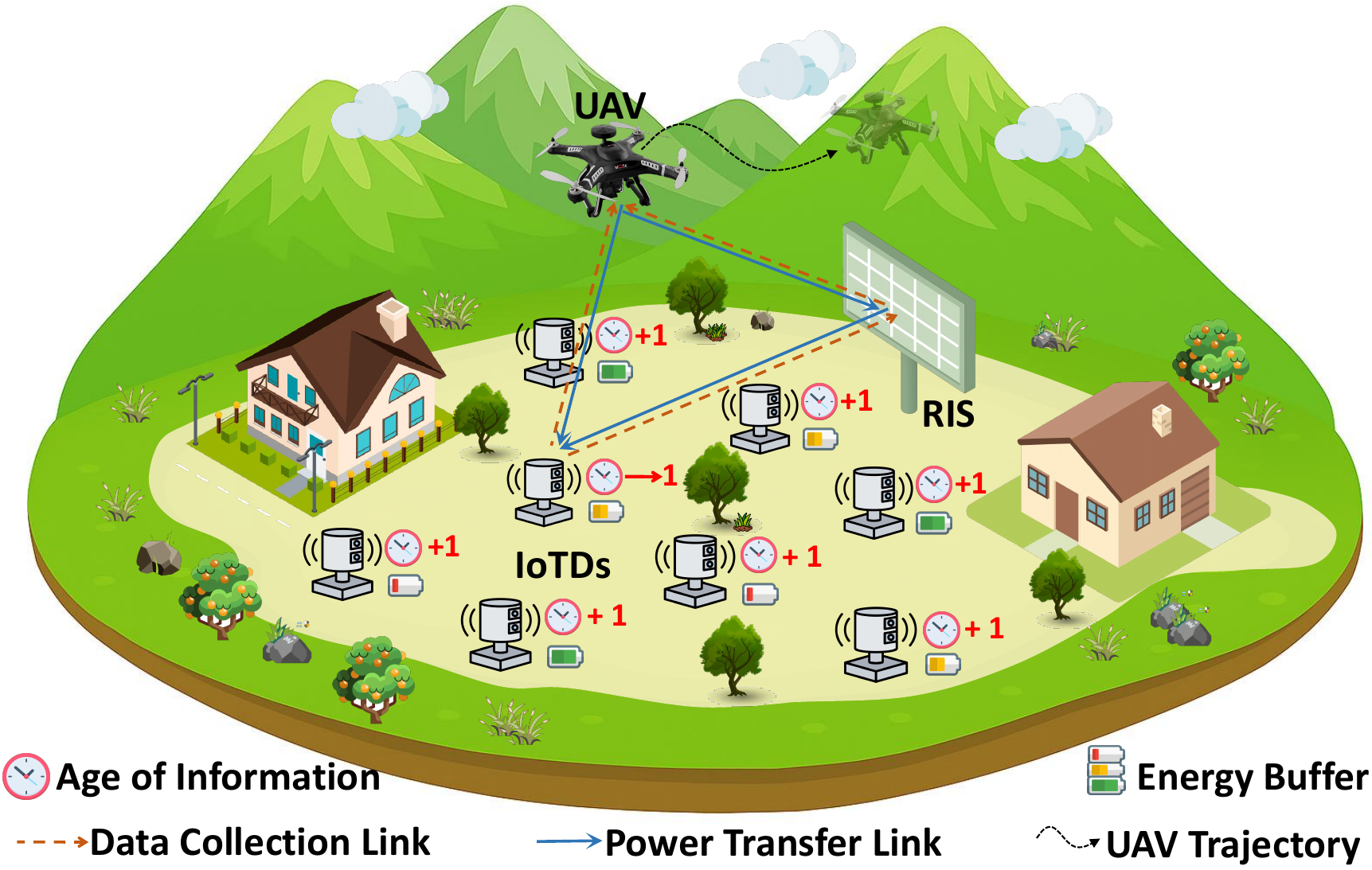}
\caption{RIS-assisted UAV-enabled data collection and wireless power transfer system in the LAWN. First, the UAV utilizes the RF-based power transfer to charge the IoTDs for ensuring sustainable energy support of IoTDs. Then, the UAV collects data from the IoTDs by using the TDMA protocol. The deployed RIS improves the channel conditions, and thus boosting data collection efficiency and energy efficiency.}
\label{fig:SystemStruction}
\end{figure}

\begin{figure}[t]
\centering
\includegraphics[width=3.5in]{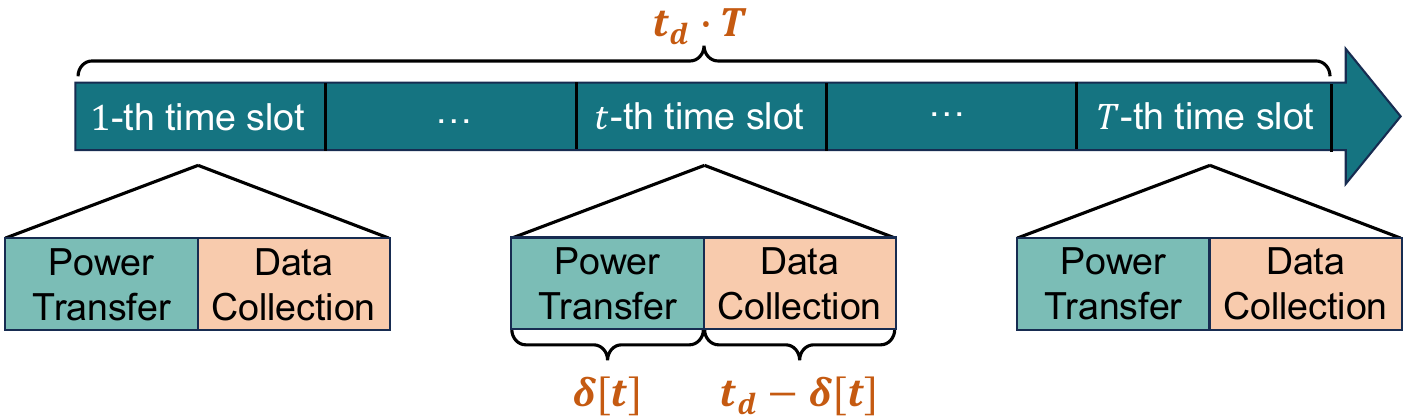}
\caption{Time allocation model. The UAV sequentially performs wireless power transfer and data collection tasks in each time slot, with durations of $\delta[t]$ and $t_{d}-\delta[t]$, respectively.}
\label{fig:Time Allocation}
\end{figure}

\subsection{System Overview}
\par As shown in Fig.~\ref{fig:SystemStruction}, we consider an RIS-assisted UAV-enabled data collection and wireless power transfer system in the LAWN, where the UAV is dispatched to collect fresh data and provide energy supply for a set of low-power IoTDs represented as $\mathcal{N} \triangleq \{1,\ldots,n,\ldots, N\}$, deployed across a target area to monitor the environment. We assume that IoTDs are located in areas with severe path loss and high attenuation due to obstacles. In this case, deploying an RIS helps enhance channel conditions while improving overall transmission quality of the system. Moreover, we assume the RIS is structured as an $M_{\rm r}\times M_{\rm c}$ element array, where the spacing is $d_{\rm r}$ for rows and $d_{\rm c}$ for columns.

\par As shown in Fig.~\ref{fig:Time Allocation}, the system operates in two phases, \textit{i.e.}, wireless power transfer phase and data collection phase. Specifically, considering the energy constraints of low-power IoTDs, the UAV wirelessly charges them using RF-based power transfer in the first phase. Then, the selected IoTD uploads its sensing data to the UAV using the energy previously harvested and stored in its buffer in the second phase. In this case, we discretize the operating time of the considered system into $T$ time slots with equal length $t_{\rm d}$, denoted as $\mathcal{T} \triangleq \{1,\ldots,t,\ldots,T\}$, where the wireless power transfer phase and the data collection phase are executed sequentially within each time slot to prevent mutual interference. We define the duration allocated to the wireless power transfer phase as $\delta[t] \in [0,t_{\rm d}]$, and allocate the remaining time $t_{\rm d}-\delta[t]$ to the data collection phase in each time slot. In particular, the data collection phase is scheduled after the wireless power transfer phase so that the IoTDs can replenish energy in advance, which helps reduce the risk of transmission failures caused by insufficient energy~\cite{Liang2025}. Moreover, to further avoid interference among IoTDs, the TDMA protocol schedules the data collection such that each time slot is assigned to only one IoTD for uploading data to the UAV~\cite{Fan2023}. 

\par \textit{Remark 1:} When IoTDs are distributed over a wider area, the considered system can be expanded to the system involving multiple RISs to mitigate the severe path loss and limited IoTD transmission range, thereby improving the data collection and wireless power transfer performance of the system. In this case, when an IoTD is selected to upload data to the UAV in a given time slot, we can select the RIS closest to the IoTD to assist in data collection and wireless power transfer in this time slot.

\par \textit{Remark 2:} We can leverage a high-altitude platform (HAP) to perform the long-range high-power wireless power transfer to the UAV, which can significantly improve its endurance. It is worth noting that the integration of an HAP does not affect other components of the system or the formulated optimization problem in Section~\ref{sec:Problem Formulation and Analysis}. This is because the HAP only serves as an additional energy source for the UAV and does not involve the data collection and wireless power transfer process between the UAV and IoTDs.

\par In addition, we consider a three-dimensional (3D) Cartesian coordinate system, where the location of UAV in time slot $t$ can be denoted as $q^{\rm U}[t]=[x^{\rm U}[t],y^{\rm U}[t],H^{\rm U}]$ with the fixed flight altitude $H^{\rm U}$. Similarly, the locations of IoTD $n$ and RIS can be represented as $q_{n}^{\rm D}= [x_{n}^{\rm D},y_{n}^{\rm D},0]$ and $q^{\rm I} = [x^{\rm I},y^{\rm I},z^{\rm I}]$, respectively. For simplicity, the key parameters of the system model are listed in Table~\ref{tab:Notation}.

\par In the considered system, the mobility of the UAV during data collection introduces significant system dynamics, while the limited energy of ground IoTDs leads to uncertainties in data uploading. In the following, to characterize the dynamics and uncertainties faced by the system, we model the ground-air transmission process, the IoTD energy harvest process, the energy consumption of the UAV, and the data freshness of IoTDs. 

\begin{table}[t]
    \centering
    \renewcommand{\arraystretch}{1.2}
    \caption{Major Notations}
    \label{tab:Notation}
    \begin{tabularx}{\linewidth}{lX}
    \Xhline{1.2pt} 
    Notation & Definition \\
    \Xhline{1.2pt}
    $a^{\rm x}[t]$, $a^{\rm y}[t]$ & Horizontal flight distances of the UAV in the x-axis and y-axis directions in time slot $t$ \\
    $A_{n}[t]$, $A[t]$ & AoI of $n$-th IoTD and average AoI of all IoTDs in time slot $t$, respectively \\
    $d_{\rm r}$, $d_{\rm c}$ & Spacing of reflecting elements in each row and column, respectively \\
    $E_{n}^{\rm H}$[t] & Harvested energy of $n$-th IoTD in time slot $t$ \\
    $E^{\rm C}[t]$, $E^{\rm P}[t]$, $E^{\rm U}[t]$ & Charging energy consumption, propulsion energy consumption, and total energy consumption of the UAV in time slot $t$\\
    $H_{n}^{\rm UD}[t]$,$H^{\rm UR}[t]$,$H_{n}^{\rm RD}[t]$ & Channel vectors of UAV-IoTD $n$ link, UAV-RIS link, and RIS-IoTD $n$ link, respectively \\
    $H_{n}^{\rm URD}[t]$ & Reflecting channel vector of between the UAV and $n$-th IoTD in time slot $t$. \\
    $k_{1}$, $k_{2}$, $k_{3}$ & Path loss components of UAV-IoTD $n$ link, UAV-RIS link, and RIS-IoTD $n$ link, respectively \\
    $M_{\rm r}$, $M_{\rm c}$ & Number of reflecting elements in each row and column, respectively \\
    $n$, $N$, $\mathcal{N}$ & Index, total number, and set of IoTDs \\
    $P_{n}^{\rm D}$, $P^{U}$ & Transmit power of $n$-th IoTD and the UAV, respectively\\
    $q^{\rm U}[t]$ & Location of the UAV in time slot $t$ \\
    $q_{n}^{\rm D}$, $q^{\rm I}$ & Locations of the $n$-th IoTD and RIS \\
    $R_{n}^{\rm D}[t]$ & Achievable rate of the $n$-th IoTD \\
    $t$, $T$, $\mathcal{T}$ & Index, total number, and set of time slots \\
    $t_{\rm d}$ & Duration of each time slot \\
    $Z_{\rm min}$ & Minimum data size threshold. \\
    $\alpha_{0}$ & Channel gain at the reference distance $d_{0}$=1m \\
    $\delta[t]$ & Duration of the power transfer phase in each time slot \\
    $\sigma^{2}$ & Noise power \\
    $\Phi[t]$ & RIS reflection phase coefficient matrix in time slot $t$ \\
    \Xhline{1.2pt}
    \end{tabularx}
\end{table}

\subsection{Channel and Transmission Model}
\label{Channel Model}
\par In the RIS-assisted UAV-enabled IoT system, due to the RIS deployment, the communication channel between the UAV and IoTDs is divided into two parts that are the direct link (UAV-IoTD link) and reflecting link (UAV-RIS-IoTD link), respectively.

\subsubsection{Direct Link Channel Model}

\par Given that obstacles can induce partial signal scattering in the air-ground transmission process, we use the Rician fading model to characterize the communication link between the UAV and IoTDs, incorporating a dominant LoS component alongside scattered multipath effects. As such, the channel vector related to the UAV and the $n$-th IoTD during time slot $t$ is formulated as
\begin{eqnarray}
\label{equ_1}
H_{n}^{\rm UD}[t]&=&\sqrt{\frac{\alpha_{0}}{(d_{n}^{\rm UD}[t])^{k_{1}}}}(\sqrt{\frac{r_{n}^{\rm UD}}{r_{n}^{\rm UD}+1}} e^{\frac{-j2 \pi d_{n}^{\rm UD}[t]}{\lambda}} \nonumber \\
 &+&\sqrt{\frac{1}{r_{n}^{\rm UD}+1}} g_{n,\rm NLoS}^{\rm UD}[t]), 
\end{eqnarray}
\noindent where $\alpha_{0}$ represents the channel gain measured at the reference distance $d_{0} = 1$ m, while the Rician factor associated with the direct link is denoted by $r_{n}^{\rm UD}$. Moreover, $k_{1}$ represents the path loss exponent for the UAV-IoTD link. In addition, $\lambda$ corresponds to the carrier wavelength, and the distance between the UAV and the $n$-th IoTD in time slot $t$ is defined as $d_{n}^{\rm UD}[t] = \sqrt{\|q^{\rm U}[t] - q_{n}^{\rm D}\|^{2}}$. The random variable $g_{n,\rm NLoS}^{\rm UD}[t] \sim \mathcal{CN}(0,1)$ is drawn from complex Gaussian distribution, which models the non-line-of-sight (NLoS) scattering effect.

\subsubsection{Reflecting Link Channel Model}
\par The reflecting link includes two components, namely the UAV-RIS link and the RIS-IoTD link. Because the RIS is positioned at a high elevation and the UAV operates at considerable altitude, the UAV-RIS link is typically subject to negligible interference and attenuation during signal propagation~\cite{Wei2021}. As such, the UAV-RIS link is modeled using a deterministic LoS model. The channel vector corresponding to the UAV-RIS link in time slot $t$ is given by
\begin{equation}
\label{equ_2}
H^{\rm UR}[t]=\sqrt{\frac{\alpha_{0}}{(d^{\rm UR}[t])^{k_{2}}}} g_{LoS}^{\rm UR}[t],
\end{equation}
\noindent where $d^{\rm UR}[t]=\sqrt{\|q^{\rm U}[t]-q^{\rm I}\|}$ represents the distance between the UAV and RIS. In addition, $k_{2}$ is the path loss component for the UAV-RIS link. Moreover, $g_{\rm LoS}^{\rm UR}[t]$ is the LoS component of the UAV-RIS link, which can be given by
\begin{eqnarray}
\label{equ_3}
g_{\rm LoS}^{\rm UR}[t]&=&[1, \ldots, e^{-j 2 \pi(M_{\rm r}-1) \frac{d_{\rm r} \sin \theta^{\rm UR}[t] \cos \xi^{\rm UR}[t]}{\lambda}}]^{\rm T} \nonumber \nonumber \\
 &\otimes&{[1, \ldots, e^{-j 2 \pi(M_{\rm c}-1) \frac{d_{\rm c} \sin \theta^{\rm UR}[t] \sin \xi{\rm UR}[t]}{\lambda}}]}^{\rm T},
\end{eqnarray}
\noindent where $\theta^{\rm UR}$ and $\xi^{\rm UR}$ denote the vertical and horizontal angles-of-arrival (AoA) of the UAV-RIS link.

\par Furthermore, the Rician fading model is adopted for the RIS-IoTD link due to signal scattering. Thus, the channel vector between RIS and $n$-th IoTD in time slot $t$ can be given by
\begin{eqnarray}
\label{equ_4}
     H_{n}^{\rm RD}[t]&=&\sqrt{\frac{\alpha_{0}}{(d_{n}^{\rm RD})^{k_{3}}}}(\sqrt{\frac{r_{n}^{\rm RD}}{r_{n}^{\rm RD}+1}} g_{n,\rm LoS}^{\rm RD}[t] \nonumber \\
     &+&\sqrt{\frac{1}{r_{n}^{\rm RD}+1}} g_{n,\rm NLoS}^{\rm RD}[t]),
\end{eqnarray}
\noindent where $r_{n}^{\rm RD}$ denotes the Rician factor associated with the RIS-IoTD link, and the distance between the RIS and $n$-th IoTD is given by $d_{n}^{\rm RD} = \sqrt{\|q^{\rm I} - q_{n}^{\rm D}\|^{2}}$. The parameter $k_{3}$ characterizes the path loss for this link. Moreover, the NLoS component follows a complex Gaussian distribution as $g_{n,\rm NLoS}^{\rm RD}[t] \sim \mathcal{CN}(0, \mathbf{I}_{M_{\rm r}M_{\rm c}})$. In addition, $g_{n,\rm LoS}^{\rm RD}[t]$ is the LoS component, which is represented as
\begin{eqnarray}
\label{equ_5}
g_{n,\rm LoS}^{\rm RD}[t]&=&[1, \ldots, e^{-j 2 \pi(M_{\rm r}-1) \frac{d_{\rm r} \sin \theta^{\rm RD} \cos \xi^{\rm RD}}{\lambda}}]^{\rm T} \nonumber \\ 
 &\otimes&{[1, \ldots, e^{-j 2 \pi(M_{\rm c}-1) \frac{d_{\rm c} \sin \theta^{\rm RD} \sin \xi^{\rm RD}}{\lambda}}]^{\rm T}},
\end{eqnarray}
\noindent where $\theta^{\rm RD}$ and $\xi^{\rm RD}$ denote the vertical and horizontal angles-of-departure (AoD) of the signal. 

\subsubsection{Transmission Model}
\par In time slot $t$, the RIS reflection phase coefficient matrix can be expressed as
\begin{equation}
    \label{equ_6}
    \Phi[t] = \rm{diag}(\phi[t]), 
\end{equation}
\noindent where $\phi[t]=[e^{j \phi^{1,1}[t]}, \ldots, e^{j \phi^{m_{\mathrm{r}}, m_{\mathrm{c}}}[t]}, \ldots, e^{j \phi^{M_{\mathrm{r}}, M_{\mathrm{c}}}[t]}]^{\mathrm{T}}$. Moreover, constrained by hardware capabilities, the phase shift values can only be selected from a finite set. Specifically, the phase shift of the $(m_{\rm r}, m_{\rm c})$-th reflecting element in time slot $t$ satisfies $\phi^{m_{\rm r},m_{\rm c}}[t] \in \left\{0, \frac{2\pi \cdot 1}{2^b}, \ldots, \frac{2\pi \cdot (2^b - 1)}{2^b} \right\}$, where $b$ denotes the number of quantization bits. Then, the achievable rate of the $n$-th IoTD in time slot $t$ is denoted as
\begin{equation}
\label{equ_7}
    R_{n}^{\rm D}[t] = \log_2\left(1+\frac{P_{n}^{\rm D}{|H_{n}^{UD}[t]+H_{n}^{\rm URD}[t]|}^2}{\sigma^2}\right),
\end{equation}
\noindent where $H_{n}^{\rm URD} = (H_{n}^{\rm RG}[t])^{\rm T}\Phi[t]H^{\rm UR}[t]$ is the reflecting channel between the UAV and the $n$-th IoTD with the assistance of the RIS in time slot $t$. Moreover, $P_{n}^{\rm D}$ is the transmit power of the $n$-th IoTD and $\sigma^{2}$ represents the noise power.

\subsection{Energy Model}
\par The energy model of the system consists of two parts that are the energy harvested by the IoTDs and energy consumed by the UAV.

\subsubsection{IoTD Energy Storage}
\par In the wireless power transfer phase, the UAV transmits power to the IoTDs for charging. Specifically, the widely adopted energy model~\cite{Nguyen2022} is utilized to describe the harvested energy of $n$-th IoTD from the UAV in time slot $t$, \textit{i.e.},
\begin{eqnarray}
\label{equ_8}
    E_{n}^{\rm H}[t]=\delta[t] \eta P_{\rm U} |H_{n}^{\rm UD}[t]+H_{n}^{\rm URD}[t]|^{2},
\end{eqnarray}
\noindent where $\delta[t]$ represents the charging time, and $P_{\rm U}$ stands for the transmit power of the UAV. Moreover, the efficiency of wireless power transfer is denoted by $\eta \in (0,1)$. In addition, each IoTD is subject to a maximum energy storage capacity represented by $E_{\rm max}$. Accordingly, the energy stored in the energy buffer of the $n$-th IoTD in time slot $t+1$ can be denoted as
\begin{equation}
    \label{equ_11}
     E_{n}^{\rm B}[t+1]= \left\{ 
    \begin{array}{ll}
        E_{\rm max}, & \text{if} \quad E_{n}^{\rm B}[t]+E_{n}^{\rm H}[t] \geq E_{\rm max},  \\
        E_{n}^{\rm B}[t]+E_{n}^{\rm H}[t], & \text{otherwise},
    \end{array}
    \right.
\end{equation}
\noindent where $E_{n}^{\rm B}[t]$ represents the residual energy of the energy buffer in the $n$-th IoTD at time slot $t$.

\par \textit{Remark 3:} Similar to the existing works~\cite{Liu2025,Liu2021,Oubbati2021}, we adopt the simple linear energy harvesting model. Notably, the adopted energy harvesting model can be flexibly replaced by other models, such as the non-linear model, which does not compromise the performance of the proposed approach in Section~\ref{sec:Proposed Soution}. This is because adopting different energy harvesting models only changes the mathematical expression of the energy-related constraint in the AoI metric show in Eq.~(\ref{equ_16}) and do not fundamentally change the overall formulated optimization problem show in Eq.~(\ref{equ_18}), thereby remaining the applicability and effectiveness of the proposed approach. Moreover, the proposed approach is based on the DRL algorithm, inherently possessing strong adaptability and real-time decision-making capability. As such, even when different energy harvesting models are considered, the DRL-based approach can be retrained to accommodate the new energy harvesting characteristics and still achieve effective decision-making.

\subsubsection{UAV Energy Consumption}
\par The energy consumption of the UAV is attributed to two main components, $\textit{i.e.}$, charging energy consumption and propulsion energy consumption. In terms of energy consumption caused by UAV charging IoTDs, it can be denoted as
\begin{equation}
\label{equ_12}
    E^{\rm C}[t] = P_{\rm U}\delta[t].
\end{equation}
\par In respect to the UAV propulsion energy consumption, we adopt a classical energy model~\cite{Zeng2019} to present it, which is given by
\begin{align}
\label{equ_13}
E^{\rm P}[t]&=\left(P_{\rm s}\left(1+3\left(\frac{v^{\rm h}[t]}{U_{\rm r}}\right)^2\right) + P_{\rm m}\left( \sqrt{1+\frac{1}{4}\left(\frac{v^{\rm h}[t]}{V_{\rm h}}\right)^4} \right. \right.
\nonumber \\ &\left.\left.
-\frac{1}{2}\left(\frac{v^{\rm h}[t]}{V_{\rm h}}\right)^2\right)^\frac{1}{2} +\frac{1}{2}d_0\rho_{\rm a}zG\left(v^{\rm h}[t]\right)^3\right)t_{\rm d},
\end{align}
\noindent where $U_{\rm r}$ and $V_{\rm h}$ denote the tip speed of the rotor blade and the mean rotor-induced velocity when hovering, respectively. Moreover, $d_0$, $\rho_{\rm a}$, $z$, and $G$ represent the main body drag ratio, air density, rotor solidity, and rotor disc area, respectively. In addition, $P_{\rm s}$ and $P_{\rm m}$ are two fixed constants. Furthermore, $t_{\rm d}$ is the time duration of each time slot. Additionally, $v^{\rm h}[t]$ denotes the horizontal velocity in time slot $t$, \textit{i.e.},
\begin{equation}
\label{equ_14}
v^{\rm h}[t] = \frac{\sqrt{(a^{\rm x}[t])^{2}+(a^{\rm y}[t])^{2}}}{t_{\rm d}},
\end{equation}
\noindent where $a^{\rm x}[t] \in [-a_{\rm max}^{\rm x},a_{\rm max}^{\rm x}]$ and $a^{\rm y}[t] \in [-a_{\rm max}^{\rm y},a_{\rm max}^{\rm y}]$ denote the horizontal flight distances of the UAV in the $x$-axis and $y$-axis directions in time slot $t$, respectively.

\par Thus, the total energy consumption of UAV within $T$ time slots can be denoted as
\begin{equation}
\label{equ_15}
    E^{\rm U}[t] = E^{\rm P}[t]+E^{\rm C}[t].
\end{equation}

\subsection{AoI Model}

\par Data freshness is typically quantified using the AoI metric. It is assumed that every IoTD performs information sampling once per time slot, and each IoTD maintains a single-packet queue, where outdated data is replaced by newly received data. Therefore, the AoI for the $n$-th IoTD in time slot $t+1$~\cite{Oubbati2021} is expressed as
\begin{equation}
    \label{equ_16}
    A_{n}[t+1] = \left \{
    \begin{array}{ll}
        1, &  \text{if} \ \alpha_{n}[t] = 1 \; \text{and} \\ & \;(t_{\rm d}-\delta[t])R_{n}^{\rm D}[t] \geq Z_{\rm min} \;\text{and} \\
        & \; (t_{\rm d}-\delta[t])P_{n}^{\rm D} \leq E_{n}^{\rm B}[t],\\
        A_{n}[t] + 1, & \text{otherwise},
    \end{array}
    \right.
\end{equation}
\noindent where $\alpha_{n}[t]\in\{0,1\}$ represents the IoTD scheduling, and $\alpha_{n}[t]=1$ if the IoTD $n$ is selected to upload data to the UAV, $\alpha_{n}[t]=0$, otherwise. Moreover, let$Z_{\rm min}$ be the minimum data size in bits required for reliable recovery of the data collected by the UAV~\cite{Oubbati2021}. Therefore, the AoI metric shown in Eq.~(\ref{equ_16}) indicates that the AoI of the $n$-th IoTD can be reset to $1$ only when the transmission bits exceed the threshold $z_{\rm min}$ and the residual energy can support this transmission; otherwise, the AoI of the $n$-th IoTD increases by $1$.

\par Then, the average AoI of all IoTDs in time slot $t$ is given by
\begin{equation}
    \label{equ_17}
    A[t] = \frac{1}{N}\sum\nolimits_{n=1}^{N}A_{n}[t].
\end{equation}

\section{Problem Formulation and Analysis}
\label{sec:Problem Formulation and Analysis}

\par In this section, we first formulate the optimization problem and then analyze its characteristics.

\subsection{Problem Formulation}
\par First, we define the optimization variables as follows: \textit{(i)} $\Phi = \{ \Phi[t]\ | \ \forall t\in \mathcal{T}\}$ denotes the RIS phase shifts. \textit{(ii)} $\mathcal{Q}=\{q^{U}[t]\ |\ \forall t \in \mathcal{T}\}$ represents the UAV trajectory. \textit{(iii)} $\Delta = \{ \delta[t]\ | \forall t \in \mathcal{T}\}$ is the charging time for the IoTDs. \textit{(iv)} $\alpha = \{\alpha_{n}[t]\ | \ \forall n \in\mathcal{N}, \ t\in \mathcal{T}\}$ represents the IoTD scheduling.
\par Then, we present the optimization objectives as follows:

\par \textit{Objective 1}: Information freshness is crucial for IoT applications, as it directly impacts the quality of decision-making. In particular, outdated information can lead to significant losses. Therefore, our first optimization objective is to reduce the AoI, which is denoted by
\begin{equation}
    \label{equ_optimization_objective1}
    f_{1}(\Phi,\mathcal{Q},\Delta,\alpha) = \sum_{t=1}^{T}A[t].
\end{equation}

\par \textit{Objective 2}: Although UAV mobility enables flexible data collection and power transfer, its limited energy restricts further enhancements in system performance. Therefore, the second optimization objective aims at reducing UAV energy consumption to enhance system sustainability, which is represented as
\begin{equation}
    \label{equ_optimization_objective2}
    f_{2}(\Delta, \mathcal{Q}) = \sum_{t=1}^{T}E^{\rm U}[t].
\end{equation}

\par Accordingly, based on the abovementioned optimization variables and objectives, the optimization problem is formulated as follows:
\begin{subequations}
\label{equ_18}
\begin{align}
\mathcal{P}1: \ &\min_{\Phi,\mathcal{Q}, \Delta, \alpha}(f_{1}, f_{2}), \label{Za}\\
\text{s.t.}: \
& X_{\rm min} \leq x^{\rm U}[t] \leq X_{\rm max}, \quad \forall t, \label{Zb}\\
&Y_{\rm min} \leq y^{\rm U}[t] \leq Y_{\rm max}, \quad \forall t,  \label{Zc}\\
&\| a^{\rm x}[t] \| \leq x_{\rm max}, \quad \forall t,  \label{Zd}\\
&\| a^{\rm y}[t] \| \leq y_{\rm max}, \quad \forall t, \label{Ze}\\
&\phi^{m_r,m_c}[t] \in \{0, \frac{2\pi \ldots 1}{2^{b}}, \cdots, \frac{2\pi\cdot(2^{b}-1)}{2^{b}}\}, \forall t, \label{Zf} \\
&0 \leq \delta[t] \leq t_{d}, \quad \forall t, \label{Zg} \\
&\alpha_{n}[t] \in \{0, \ 1\}, \quad \forall n, \ t, \label{Zh} \\
&\sum\nolimits_{n=1}^{N}\alpha_{n}[t]=1, \quad \forall t, \label{Zi}
\end{align}
\end{subequations}
\noindent where constraints (\ref{Zb}) and (\ref{Zc}) ensure the UAV stays inside the specified target area, while its flight distance per time slot is limited by (\ref{Zd}) and (\ref{Ze}). Moreover, constraint (\ref{Zf}) represents the RIS phase shifts selected from a predefined set. The charging time allocation should satisfy the constraint (\ref{Zg}). In addition, constraint (\ref{Zh}) indicates that the IoTD scheduling variable can only select 0 or 1. Furthermore, constraint (\ref{Zi}) ensures that only one IoTD is assigned to each time slot.


\subsection{Problem Analysis}

\par It is observed that the formulated optimization problem above possesses the following characteristics.

\par \textit{\uline{Non-convex}}: First, the optimization variables in the optimization problem are highly coupled due to their complex interdependencies. Second, the RIS phase shift values are restricted to a predefined discrete set, which results in a non-convex feasible region for the optimization problem. Finally, the optimization problem involves both continuous variables, such as UAV trajectory and charging duration, and a binary discrete variable corresponding to IoTD scheduling, making it a mixed-integer problem with inherent non-convex characteristics. As such, the optimization problem exhibits non-convex characteristics.
\par \textit{\uline{NP-hard}}: Maximizing the received power in a single-user scenario by optimizing the RIS phase shifts can be considered a quadratically constrained quadratic program (QCQP) problem, which has been proven to be NP-hard~\cite{Wu2019a}. For simplicity, we simplify the original optimization problem shown in Eq.~(\ref{equ_18}). Specifically, we only consider minimizing the AoI by optimizing the RIS phase shifts while fixing the UAV trajectory, charging time, and IoTD scheduling. Notably, the AoI metric depends on both the transmission rate and the harvested energy, which are both positively correlated with the received power of the UAV and IoTDs. In this case, the optimization objective in the simplified optimization problem is essentially to maximize the received power. Since maximizing the received power in a single-user scenario by optimizing RIS phase shifts has been proven to be NP-hard, the simplified optimization problem is also NP-hard. Given that the optimization problem shown in Eq.~(\ref{equ_18}) is inherently more complicated than the simplified version, it also exhibits NP-hard characteristics.

\par \textit{\uline{Dynamics}}: The flight speed and direction of the UAV should be dynamically adapted according to the freshness of data from ground IoTDs in the considered system. At the same time, RIS phase shifts should be reconfigured according to the UAV location and the IoTD scheduling to adjust the passive beamforming direction. Therefore, the channel conditions of the system change in real time with the changing RIS phase shifts, UAV trajectory, and IoTD scheduling. In this case, the two optimization objectives in the optimization problem evolve over time. As such, the formulated optimization problem is essentially a dynamic problem.

\par \textit{\uline{Long-term Optimization}}: Since the formulated optimization problem is dynamic and we aim to optimize the overall performance across $T$ time slots, the optimal solution in any single time slot cannot represent the final optimal solution over $T$ time slots. As such, the formulated problem exhibits long-term characteristics, which requires a balance between short-term objectives and long-term objectives.

\par Notably, conventional optimization algorithms are not suitable for solving the complex and dynamic optimization problem. Specifically, conventional optimization methods, such as convex optimization and evolutionary algorithms, typically rely on precise and complete prior knowledge. However, in the considered dynamic wireless scenario, obtaining high-precision prior knowledge is typically infeasible. Moreover, most conventional algorithms are designed for one-time offline solutions, which makes it challenging for them to respond to dynamic system conditions. Therefore, the adaptability and robustness of conventional algorithms in such a highly dynamic scenario are significantly limited.

\par In this case, DRL algorithms can be regarded as promising solutions to the formulated optimization problem. \textit{First}, DRL algorithms allow the agent to interact with the environment and continuously adjust its policy based on received rewards, without requiring precise prior knowledge. \textit{Second}, DRL algorithms are capable of making real-time decisions according to the environment state, which enables a fast response to dynamic environment changes. \textit{Finally}, DRL algorithms can achieve a balance between short-term and long-term objectives during the long-term operation of the system by adjusting the discount factor. As such, we propose a DRL-based approach to deal with the optimization problem shown in Eq.~(\ref{equ_18}).


\section{Proposed Solution}
\label{sec:Proposed Soution}

\par In this section, we first reformulate the optimization problem as an MDP for the facility of applying DRL. Subsequently, we analyze the challenges faced by conventional DRL algorithms to deal with the formulated optimization problem. Based on this, we propose an improved DRL-based approach with more powerful exploration capability and learning efficiency.

\subsection{MDP Formulation}
\par In DRL, the dynamics of agent-environment interaction are generally represented through the MDP framework. Formally, an MDP is typically modeled by the five-tuple $<\mathcal{S},\mathcal{A},\mathcal{R},\mathcal{P}_{\rm sa},\gamma>$. Here, the state space $\mathcal{S}$ defines the set of all possible situations in the interaction process, while the action space $\mathcal{A}$ represents the set of all feasible actions of the agent. The reward function is denoted by $\mathcal{R}$, while $\mathcal{P}_{\rm sa}$ characterizes the state transition dynamics. Moreover, $\gamma$ is the discount factor, which controls how much future rewards contribute to the overall return. Among these elements, $\mathcal{S}$, $\mathcal{A}$, and $\mathcal{R}$ play particularly important roles, and thus they are explained in detail in the following.

\subsubsection{State Space}
\par Since the agent can observe the conditions of the UAV and IoTDs, the state $s[t] \in \mathcal{S}$ is expressed as
\begin{equation}
\label{equ_23}
    s[t]=\{q^{U}[t], A^{\rm I}[t], E^{\rm B}[t]\},
\end{equation}
\noindent where $q^{U}[t]$ is the location of the UAV in time slot $t$. Moreover, $A^{\rm I}[t]=(A_{1}[t],\ldots,A_{N}[t])$ and $E^{\rm B}[t]=(E_{1}^{\rm B}[t],\ldots,E_{N}^{\rm B}[t])$ represent the AoI and the residual energy of all IoTDs in time slot $t$.

\subsubsection{Action Space}
\par In time slot $t$, the action $a[t] \in \mathcal{A}$ is selected by the agent according to the state $s_{t}$, $\textit{i.e.,}$
\begin{equation}
\label{equ_24}
    a[t]=\{a^{\rm x}[t], a^{\rm y}[t], \Phi[t],\delta[t],\alpha[t]\},
\end{equation}
\noindent where $a^{\rm x}[t]$ and $a^{y}[t]$ denote the flying distances of the UAV in $x$-axis and $y$-axis directions in time slot $t$, respectively. Moreover, $\Phi[t]$ and $\delta[t]$ represent the RIS phase shifts and charging time, respectively. In addition, $\alpha[t]=\{n\in\mathcal{N}|\alpha_{n}[t]=1\}$ represents the IoTD scheduling for uploading data in time slot $t$.

\subsubsection{Reward Function}
\par To ensure that the DRL agent learns effectively, the reward function is tailored to match the desired optimization objectives. In this case, the reward $r[t] \in \mathcal{R}$ is given by
\begin{equation}
\label{equ_25}
    r[t] = r^{\rm A}[t]+r^{\rm E}[t]+r^{\rm P}[t],
\end{equation}
\noindent where $r^{\rm A}[t]=-\sum_{n=1}^{N}(A_{n}[t-1]-A_{n}[t])$ and $r^{\rm E}[t] = -\frac{E^{\rm U}[t]}{\omega}$ are related to two optimization objectives, respectively. We introduce the scaling factor $\omega$ to balance their magnitudes, ensuring that the agent does not become biased toward optimizing a single optimization objective. Moreover, constraints (\ref{Zb}) and (\ref{Zc}) should be satisfied when the agent chooses actions. As such, the penalty $r^{\rm P}[t]$ is designed to restrain the behavior of the UAV, which is given by
\begin{equation}
\label{equ_26}
r^{\rm P}[t] = \left \{
\begin{array}{ll}
    0, &  \text{if} \ x^{\rm U}[t] \in [X_{\rm min},X_{\rm max}] \ \text{and} \\
    & y^{\rm U}[t] \in [Y_{\rm min},Y_{\rm max}], \\
    \mathbf{r_{1}}, & \text{otherwise}
\end{array}
\right.
\end{equation}
\noindent where $\mathbf{r_{1}}$ is the negative feedback given to the agent when the UAV operates outside the target area.

\subsection{The Limitations of Conventional DRL Algorithms}

\par Based on the MDP framework above, we can find that there are several inherent challenges in applying conventional DRL algorithms to solve the optimization problem.

\begin{itemize}
    \item \textit{High-Dimensional Action Space Challenge}. As can be seen, the action space in Eq.~(\ref{equ_24}) includes the RIS phase shifts. However,Since an RIS is composed of numerous reflecting elements, the configuration space of its phase shift adjustments becomes extremely high-dimensional, leading to a substantial expansion of the DRL action space. As such, directly applying conventional DRL algorithms to handle the large-dimensional action space may result in slow convergence and lower solution accuracy.
    \item \textit{Hybrid Action Space Challenge}. The action space includes both continuous variables (\textit{i.e.}, UAV trajectory and charging time allocation) and discrete variables (\textit{i.e.}, RIS phase shifts and IoTD scheduling), which makes the action space demonstrate the hybrid characteristic. However, most existing DRL algorithms are only capable of effectively handling a single type of action space. In this case, discretizing continuous actions and approximating discrete actions with continuous representations have become two commonly used methods for applying conventional DRL to hybrid action spaces. Nevertheless, both methods inevitably introduce approximation errors, which can significantly reduce the accuracy of DRL algorithms.
\end{itemize}


\par Accordingly, our main focus is to reduce the dimensionality of the action space and effectively optimize hybrid actions. To this end, we propose the AO-IPDQN approach. Specifically, we first employ an AO-based action space dimensionality reduction method to reduce the difficulty of action learning for DRL. Then, we propose an IPDQN method with improved exploration capability and learning efficiency, which integrates the core optimization mechanisms of the deep Q-network (DQN) and deep deterministic policy gradient (DDPG) to achieve joint optimization of continuous and discrete actions. In this way, the proposed AO-IPDQN approach effectively addresses the challenge of high-dimensional hybrid action spaces.

\subsection{AO-IPDQN Approach}

\par In this part, we provide detailed description of the proposed AO-IPDQN approach.


\subsubsection{AO-based Action Space Dimensionality Reduction Method}
\par With the given UAV trajectory, charging time allocation, and IoTD scheduling, the original problem $\mathcal{P}$ is reformulated by
\begin{subequations}
\begin{align}
\mathcal{P}2: \ & \min_{\Phi}\ A, \label{Z36_a}\\
\text{s.t.}: \ &
\phi^{m_r,m_c}[t] \in \{0, \frac{2\pi\cdot1}{2^{b}}, \cdots, \frac{2\pi\cdot(2^{b}-1)}{2^{b}}\}, \forall t. \label{Z36_b}
\end{align}
\end{subequations}
\par To ensure that the UAV is able to effectively collect data from the scheduled IoTD for improving the data freshness of the system, it is essential to optimize the harvested energy and transmission rate during the wireless power transfer phase and data collection phase, respectively. Thus, we employ a simple but robust AO-RIS method shown in Algorithm~\ref{Algorithm3} to configure the RIS phase shifts~\cite{Fan2023}, which is detailed as follows. 

\begin{itemize}
    \item \textit{Step 1:} The phase shifts of all RIS elements are randomly configured from the discrete phase-shift set $\phi_{\rm s}$.

    \item \textit{Step 2:} We sequentially optimize each RIS element while keeping the others fixed. In detail, for each element, the phase shift is configured by selecting the value from $\phi_{\rm s}$ that yields the maximum charging energy during the wireless power transfer phase or the highest achievable rate during the data collection phase.

    \item \textit{Step 3:} The \textit{Steps} 1 and 2 are repeated within each time slot until the maximum iteration limit is reached.
    
\end{itemize}

\par Therefore, after the abovementioned AO-RIS method optimizes the RIS phase shifts, the action space only includes the UAV trajectory, charging time allocation, and IoTD scheduling. By reducing the dimension of the action space in Eq.~(\ref{equ_24}) from $(3 + M_{\rm r}\times M_{\rm c}+N)$ to $(3+N)$, the training process of the DRL algorithm becomes significantly less complex.

\begin{algorithm}[t]
    \SetAlgoLined 
    \caption{AO-RIS}
        \label{Algorithm3}
    \LinesNumbered
	\KwIn{UAV location $q^{\rm U}[t]$, charging time allocation $\delta[t]$ and IoTD scheduling $\alpha[t]$;}
	\KwOut{Phase shifts of RIS $\Phi[t]$;}
	Randomly initialize the phase shifts from $\phi_{\rm s}  \in \{0, \frac{2\pi\cdot1}{2^{b}}, \cdots, \frac{2\pi\cdot(2^{b}-1)}{2^{b}}\}$;\\
    Initialize the maximum number of iterations $Iter_{\rm max}$;\\
    \For{i = \rm {1 to} $Iter_{\rm max}$}{
        \For{r = \rm {1 to} $M_{\rm r}$}{
            \For{c = \rm {1 to} $M_{\rm c}$}{
                $\phi^{r',c'}[t]$ is fixed, $\forall r' \not = r, c' \not = c$;\\
                \If {wireless power transfer phase}{
                Set $\phi^{r,c}[t]$ when $E_{n}^{\rm H}[t]$ is maximized;}
                \Else{
                Set $\phi^{r,c}[t]$ when $R_{n}^{\rm D}[t]$ is maximized;
                }
            }
        }
    }
\end{algorithm}

\subsubsection{IPDQN-based Hybrid Action Optimization Method}

\par In this part, we begin by providing the basic principles of the conventional parameterized deep Q-network (PDQN) method~\cite{Xiong2018}, then analyze its limitations, and finally present the IPDQN method with corresponding improvements.

\par \textit{\textbf{1) Conventional PDQN Method}}: To better represent the structural characteristics of the hybrid action space, the action space shown in Eq.~(\ref{equ_24}) is redefined as
\begin{equation}
    \label{equ_27}
    \mathcal{A} = \{(d,c_{d}) |c_{d} \in \mathcal{C}_{d}, \ \forall d \in \mathcal{D}\},
    \end{equation}
\noindent where $d \in \mathcal{D}$ denotes a discrete decision corresponding to IoTD scheduling. For each $d$, the set $\mathcal{C}_d$ contains the associated continuous variables $c_d$, which include the UAV trajectory and the allocated charging time.

\par An action-value function $Q(s, d, c_d)$ is utilized to assess the selected action by the agent under the current observed state. Accordingly, the Bellman equation can be given by
\begin{eqnarray}
\label{equ_28}
     &Q(s[t],d[t],c_{d}[t])= \mathop{\mathbb{E}}\limits_{r[t],s[t+1]}[r[t]+\gamma \mathop{\max}\limits_{d\in \mathcal{D}}  \nonumber \mathop{\sup}\limits_{c_{d} \in \mathcal{C}_{d}} \\ &Q(s[t+1],d,c_{d})|s[t]=s,a[t]=(d[t],c_{d}[t])],
\end{eqnarray}
\noindent where $\gamma$ denotes the discount factor. Given a specific state $s$, the optimal continuous action corresponding to each discrete action $d \in \mathcal{D}$ is obtained by solving $c_{d}^{*} = \arg\sup_{c_{d} \in \mathcal{C}_{d}} Q(s, d, c_d)$. For simplicity, this optimal action can be represented as a mapping from the state space to the continuous action space, written as $c_d^{Q}: \mathcal{S} \rightarrow \mathcal{C}_d$.

\par Note that the largest $Q(s[t+1],d,c_{d}^{*})$ can be obtained when the optimal continuous action $c_{d}^{*}$ is determined~\cite{Xiong2018}. However, it is intractable to take supremum over the continuous space. Thus, a deep neural network (DNN) is adopted to serve as an approximation of $c_{d}^{Q}$, which can be formally expressed as
\begin{equation}
    \label{equ_29}
    c_{d}(s;\theta_{\rm P}) \approx c_{d}^{Q}(s),
\end{equation}
\noindent where $\theta_{\rm P}$ is the parameter of the policy network.

\par Likewise, we also use the DNN to approximate the value network. In this case, after obtaining the continuous action $c_{d}$, the corresponding optimal discrete action $d^{*}$ is determined as follows: 
\begin{equation}
    \label{equ_30}
    d^{*} = \arg \max \limits_{d \in \mathcal{D}} Q(s,d,c_{d};\theta_{\rm Q}),
\end{equation}
\noindent where $Q(\cdot;\theta_{\rm Q})$ is the value network to approximate $Q(s,d,c_{d})$. Moreover, $\theta_{\rm Q}$ is the parameter of the network. In addition, two target networks are applied to improve the learning stability, $\textit{i.e.}$, target policy network $c_{d}(s;\theta_{\rm P'})$ with parameter $\theta_{\rm P}'$ and target value network $Q(s,d,c_{d};\theta_{\rm Q'})$ with parameter $\theta_{\rm Q'}$.

\par In order to learn from historical data to improve training efficiency, the conventional PDQN method uses the experience replay buffer $\mathcal{B}$ to store the experience tuples $(s[t], a[t], r[t], s[t+1])$ generated during the training phase. Moreover, during training, a mini-batch of $B$ experience samples is uniformly drawn from the replay buffer, and the samples are then used for updating network parameters.

\par As such, when the parameters $\theta_{\rm P'}$ and $\theta_{\rm Q'}$ are fixed, the loss function of the value network can be expressed as
\begin{equation}
\label{equ_34-1}
    L(\theta_{Q}) = \frac{1}{|B|} \sum \limits_{b \in B} \left[Q\left(s[b],d[b],c_{d}[b];\theta_{\rm Q}\right)-y[b]\right]^{2},
\end{equation}
\noindent where $y[b]=r[b]+\gamma\max_{d \in \mathcal{D}}Q(s[b+1],d,c_{d}(s[b+1];\theta_{\rm P'});\theta_{\rm Q'})$ represents the target value.

\par Similarly, the loss function of the policy network is given by
\begin{equation}
\label{equ_35}
    L(\theta_{P}) = \frac{1}{|B|} \sum \limits_{b \in B} \left[-\sum_{d \in \mathcal{D}}\left(Q(s[b],d,c_{d}(s[b];\theta_{\rm P});\theta_{\rm Q})\right)\right].
\end{equation}

\par Based on the aforementioned methods, the soft-update mechanism for updating the target value and policy networks is defined as
\begin{eqnarray}
\label{equ_36}
    \theta_{\rm Q'} \leftarrow \tau \theta_{\rm Q}+(1-\tau)\theta_{\rm Q'},\\
\label{equ_37}
    \theta_{\rm P'} \leftarrow \tau \theta_{\rm P}+(1-\tau)\theta_{\rm P'}, 
\end{eqnarray}
\noindent where $\tau \in [0,1]$ represents the update weight.

\par \textit{\textbf{2) Limitations of Conventional PDQN Method}}: Although the PDQN method shows advantages in solving hybrid action space optimization, it still has inherent limitations.

\begin{itemize}
    \item \textit{Low Learning Efficiency}. The random sampling strategy adopted in conventional PDQN treats all experiences in the replay buffer with equal probability, without considering the varying learning value of different samples. This undifferentiated sampling strategy limits the algorithm ability to fully exploit critical experiences that contribute to policy improvement, potentially leading to prolonged convergence time due to frequent learning of low-value experiences~\cite{Goek2024}. As such, prioritizing high-value experiences is essential for better learning efficiency.
    \item \textit{Limited Exploration Capability}. Conventional PDQN relies on the agent exploration to discover high-quality policies. However, insufficient or inefficient exploration can drive PDQN toward suboptimal solutions at an early stage, thereby degrading overall performance. Therefore, maintaining both the diversity and effectiveness of exploration during the training phase is essential for the PDQN method.
\end{itemize}
\par To address the limitations above, we propose an IPDQN method with improved learning efficiency and more powerful exploration capability.

\begin{figure*}[h]
    \centering
    \includegraphics[width=\linewidth]{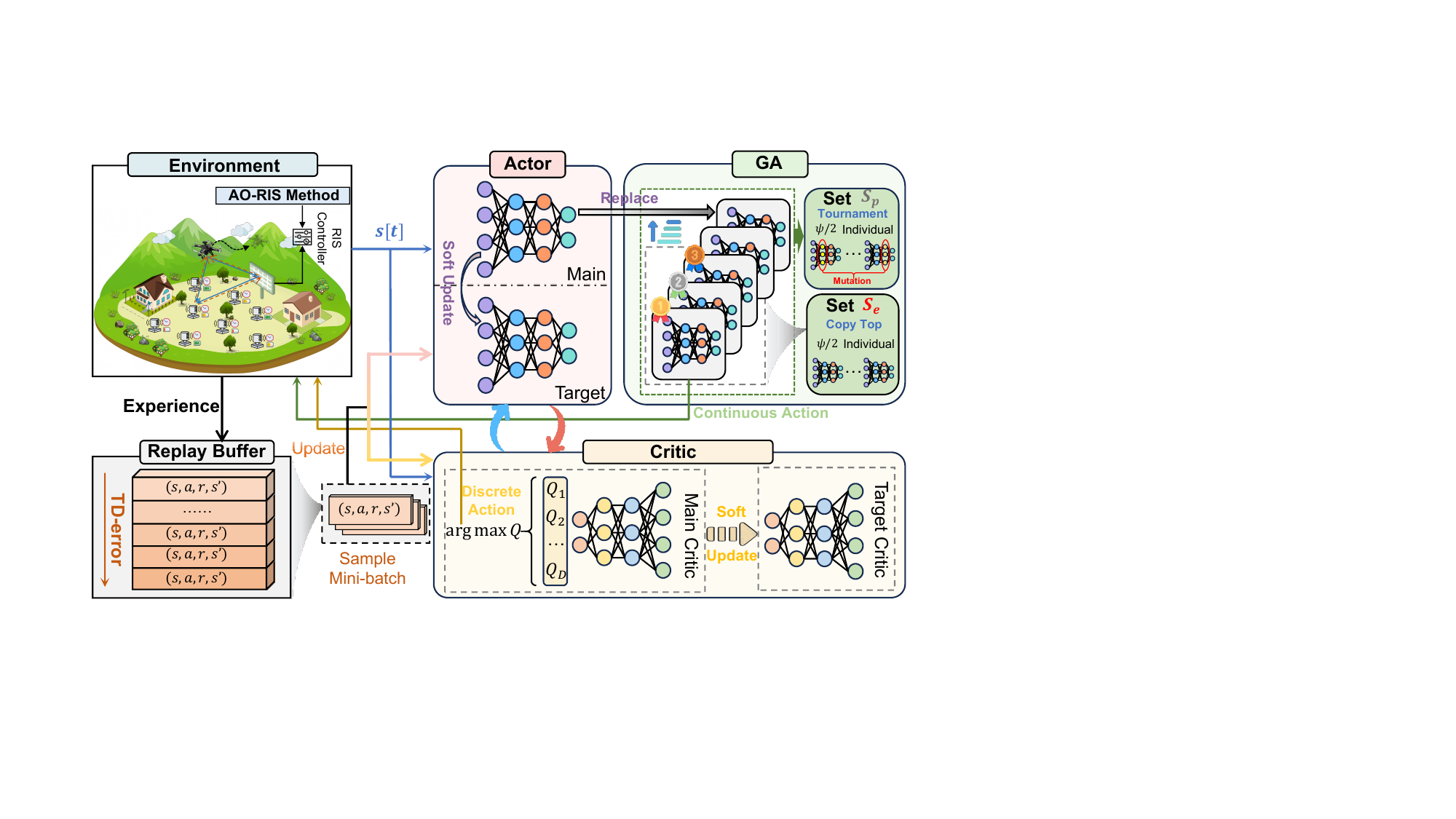}
    \caption{The framework of the proposed AO-IPDQN approach. First, the IPDQN method, which combines the PER mechanism with the GA, outputs the UAV trajectory, charging time allocation, and IoTD scheduling based on the current environment state. Subsequently, the AO-RIS method configures the RIS phase shifts based on the UAV location and IoTD scheduling. During this process, the environment evolves dynamically based on the decisions, and the resulting states, actions, and rewards are stored in the experience buffer.}
    \label{fig: approach framework}
\end{figure*}

\par \textit{\textbf{3) IPDQN Method}}: First, the proposed IPDQN method incorporates the PER mechanism to improve the efficiency of experience sampling. Second, IPDQN method leverages the evolutionary mechanism of the GA algorithm to maintain and update a population of policy networks, thereby promoting diversity in the exploration behavior of the DRL agent. In this case, the PER mechanism and GA complement each other, where the GA ensures the breadth and diversity of exploration, while the PER mechanism ensures the depth and efficiency of learning. In the following, we introduce the details of the improvements introduced in the IPDQN method.

\par $\bullet$ \textit{PER Mechanism}: Since experience samples with larger temporal-difference (TD) errors are generally more informative, PER estimates the learning value of each experience sample according to its TD error, allowing those with higher learning value to be sampled more frequently~\cite{Goek2024}. In this case, the probability of selecting the $b$-th experience in the replay buffer is given by
\begin{equation}
    \label{equ_31}
    P(b) = \frac{p_{b}^{\alpha}}{\sum \limits_{m \in B} p_{m}^{\alpha}},
\end{equation}
\noindent where $\alpha$ is a factor to determine the prioritization. Moreover, $p_{b}=|\delta_{b}|+\epsilon$ is the priority value, and $\epsilon$ represents a positive constant to avoid the condition of $p_{b}=0$. In addition, $\delta_{b}$ is the TD-error, which can be denoted as
\begin{eqnarray}
\label{equ_32}
    \delta_{b}&=&r[b]+\gamma\max_{d \in \mathcal{D}}Q(s[b+1],d,c_{d}(s[b+1];\theta_{P^{'}});\theta_{\rm Q'})- \nonumber \\
    & &Q(s[b],d[b],c_{d}[b];\theta_{\rm Q}).
\end{eqnarray}

\par Furthermore, to mitigate the effects of oscillation and divergence due to frequent sampling of high $|\delta_{b}|$ experiences, we apply importance sampling to evaluate the significance of sampled experiences, which can be represented as
\begin{equation}
    \label{equ_33}
    \omega_{b}=\frac{1}{(N_{\rm E}\cdot P(b))^{\mu}},
\end{equation}
\noindent where $N_{\rm E}$ is the size of the experience replay buffer, and $\mu$ denotes the factor to decide the importance of PER.

\par Consequently, based on experiences obtained via the PER mechanism, the loss function of the value network can be expressed as
\begin{equation}
\label{equ_34}
    L(\theta_{\rm Q}) = \frac{1}{|B|} \sum \limits_{b \in B} \omega_{b}\left[Q\left(s[b],d[b],c_{d}[b];\theta_{\rm Q}\right)-y[b]\right]^{2}.
\end{equation}

\par $\bullet$ \textit{GA-based Policy Network Optimization}: To further enhance the diversity and effectiveness of exploration, GA is employed to generate and maintain a population of candidate policy network parameter vectors. In particular, the policy population is iteratively updated through both the network update mechanism of the PDQN method and the evolutionary process of the GA. In this process, the policy population maintained by the GA facilitates more diverse exploration behaviors for the DRL agent, while the inherent redundancy characteristic within the population contributes to a more stable learning process. Meanwhile, the gradient information provided by the PDQN method improves the sample efficiency of the GA in return~\cite{Khadka2018}. Therefore, the combination of the GA and PDQN method leads to a complementary effect. The GA-based policy network optimization process is detailed as follows.
\begin{itemize}
    \item \textit{Step 1}: Initialize parameters, including the structure of the policy network $c_{d}(\cdot|{{\theta}_{\rm P}^{\rm rl}})$ and value network $Q(\cdot|{\theta}_{\rm Q})$, and the experience buffer $\mathcal{B}$. Moreover, sample $\psi$ candidate policy network parameter vectors of PDQN as the initial population of GA, $\textit{i.e.}$, ${pop}_{\rm P}=\{{\theta_{\rm P}}_{1}, \cdots, {\theta_{\rm P}}_{\psi}\}$.
    \item \textit{Step 2}:  Calculate the fitness values ${(f_{i})}_{i=1,\ldots,\psi}$ of the individuals. Note that the corresponding fitness is computed as the cumulative reward obtained during an episode in the environment. Moreover, the experience tuples $(s[t],a[t],r[t],s[t+1])$ generated in the process are stored in the experience replay buffer.
    \item \textit{Step 3}: Select $\frac{\psi}{2}$ individuals with the top fitness values to form the elite set $S_{e}$.
    \item \textit{Step 4}: Select $\frac{\psi}{2}$ individuals to form the temporary set $S_{\rm t}$ based on the tournament selection method. Specifically, in each iteration, two individuals are drawn from the population $pop_{\rm P}$, and the individual exhibiting higher fitness is incorporated into a temporary set $S_{\rm t}$.
    \item \textit{Step 5}: Utilize the crossover operation between the individuals in $S_{\rm e}$ and the individuals in $S_{\rm t}$.
    \item \textit{\textbf{Step 6}}: Perform the mutation operation for the individuals in $S_{\rm t}$. The specific approach is to add noise to individuals for perturbation according to the mutation rate $mu_{\rm p}$. Then, combine the individuals in $S_{e}$ with the individuals in $S_{t}$ to update the population ${pop}_{\rm p}$.
    \item \textit{\textbf{Step 7}}: Leverage the PER mechanism to sample a mini-batch of experience from the experience replay buffer, and Eqs.~(\ref{equ_35}) and~(\ref{equ_34}) are employed to update $\theta_{p}^{\rm rl}$ and $\theta_{\rm Q}$ with the experience sampled. 
    Then, find the individual with the lowest fitness value in $pop_{\rm p}$ and replace it with $\theta_{\rm P}$.
    \item \textit{\textbf{Step 8}}: Repeat steps 2 to 7 until the maximum iterations are reached.
\end{itemize}

\begin{algorithm}[t]
    \SetAlgoLined 
	\caption{AO-IPDQN}
        \label{Algorithm1}
    \LinesNumbered
    Initialize the weights of the value network and target value network, $\textit{i.e.}$, $\theta_{\rm Q}$, and $\theta_{\rm Q'}$;\\
    Initialize $\theta_{\rm P}^{\rm rl}$ and $\theta_{\rm P'}^{\rm rl}$ as the weights of the policy network and target policy network;\\
    Initialize experience replay buffer $\mathcal{B}$;\\
    Initialize $\psi$ policy network parameter vectors $\theta_{\rm P}$ to form population $pop_{\rm P}$;\\
    Initialize the maximum number of episodes $E_{\rm p}$ and time slot length $T$;\\
    \For{episode = \rm{1 to} $E_{\rm p}$}{
        \For{i=\rm{1 to} $\psi$}{
            Set the policy parameter ${\theta_{\rm p}}_{i}$ to policy $c_{d}$; \\
            Calculate the fitness value $f_{i}$ for policy $c_{d}(\cdot|\theta_{\text{p}_{i}})$ according to Algorithm~\ref{Alogrithm2};\\
        }
        Rank the population based on the fitness value;\\ 
        Select the top $\frac{\psi}{2}$ policy network parameter vectors to form elite set $S_{\rm e}$;\\
        Choose $\frac{\psi}{2}$ individuals through tournament selection to for temporary set $S_{\rm t}$;\\
        Use the crossover operation between $S_{\rm t}$ and $S_{\rm e}$ to generate new individuals and replace $S_{\rm t}$ with them; \\
        \For{$\theta_{\rm p} \in \rm{Set} \ S_{t}$}{
            \If{$r() \leq mu_{\rm p}$}{
                Execute mutation operation for $c_{d}(\cdot|\theta_{\rm p})$; 
            }
        }
        
        Sample a mini-batch of experiences $B$ through PER technique;\\
        Use $B$ to update the parameters $\theta_{\rm Q}$ and $\theta_{\rm P}^{\rm rl}$ according to Eqs.~(\ref{equ_34}) and~(\ref{equ_35}); \\
        Update the two target networks according to Eqs.~(\ref{equ_36}) and~(\ref{equ_37});\\

        Evaluate DRL policy network parameter vector $\theta_{\rm P}^{\rm rl}$ for fitness value $f_{\rm rl}$ according to Algorithm~\ref{Alogrithm2};\\

        Replace the weakest individual in the population with $\theta_{\rm P}^{\rm rl}$;
    }
    Return the policy parameters with the best fitness.
\end{algorithm}

\begin{algorithm}[t]
    \SetAlgoLined 
    \caption{Evaluate for Fitness}
    \label{Alogrithm2}
    \LinesNumbered
    \KwIn{Time slot length $T$, policy network $c_{d}(\cdot|\theta_{\rm P})$, and value network $Q(\cdot|\theta_{\rm Q})$;}
    \KwOut{Fitness $F$;}
    $F = 0$;\\
    \For{t = \rm{1 to} $T$}{
        Choose continuous actions $c_{d}[t] \leftarrow c_{d}(s[t];\theta_{\rm P})$;\\
        Choose discrete action $d[t] \leftarrow \arg\max_{d \in \mathcal{D}}Q(s[t],d,c_{d}[t]|\theta_{\rm Q})$; \\
        Determine the action $a[t]$ based on $\epsilon$-greedy: \\

        a[t]=$\left\{
        \begin{array}{ll}
        \rm{sample \ from \ a \ distribution, \ with \ probability \ \epsilon}, \\
        (c_{d}[t],d[t]), \rm{\ with \ probability \ 1-\epsilon}.
        \end{array}
        \right.$

        Get the phase shifts according to Algorithm~\ref{Algorithm3};\\
        Obtain a new state $s[t+1]$ and a reward $r[t]$;\\
        $F = F + r[t]$;\\
        Fill $\mathcal{B}$ with the collected experience;\\
    }
\end{algorithm}

\subsubsection{Main Flow of AO-IPDQN Approach}

\par In this part, we first present the training and execution process of the proposed AO-IPDQN approach. Subsequently, we analyze its computation complexity.

\par \textit{\textbf{1) Training and Execution}}: The pseudocode of the AO-IPDQN method is provided in Algorithm~\ref{Algorithm1}, and the overall framework is illustrated in Fig.~\ref{fig: approach framework}. In the training phase, the IPDQN method is first applied to jointly optimize the UAV trajectory, charging time allocation, and IoTD scheduling. Based on the optimization variables above, the phase shifts of all RIS reflecting elements are configured through the AO-RIS method described in Algorithm~\ref{Algorithm3}. After executing the actions generated by the IPDQN method and configuring the corresponding RIS phase shifts, the agent obtains the reward feedback and observes the new environment state. Moreover, all experiences generated during the training phase are stored in the replay buffer for subsequent neural network parameter updates. After sufficient training, the actor network exhibiting the highest fitness in the trained IPDQN method is utilized together with the AO-RIS method to make decisions according to the real-time environment state.

\par \textit{\textbf{2) Computation Complexity}}: The computation complexity of the proposed AO-IPDQN approach is divided into two parts, $\textit{i.e.}$, the complexity of the AO-RIS method and the complexity of the IPDQN method.

\par \textit{Computation Complexity of AO-RIS Method}: The complexity of the AO-RIS method is related to the number of iterations $Iter_{max}$, the size of RIS elements $M_{r}\times M_{c}$, and the length of the discrete phase shift set $2^{b}$. The computation complexity of the AO-RIS method is expressed as $\mathcal{O}(Iter_{max} \times M_{r} \times M_{c} \times 2^{b})$~\cite{Fan2023}. 
\par \textit{Computation Complexity of IPDQN Method}: The complexity of the PDQN method is mainly determined by the structure of the policy and value networks. Assume that the number of the fully-connected layers in the policy and value networks is $\mathscr{A}$ and $\mathscr{C}$, respectively. Moreover, let $\mathscr{A}_{i}$ be the number of neurons in the $i$-th layer of the policy network. Likewise, $\mathscr{C}_{j}$ denotes the number of neurons in the $j$-th layer of the value network. In addition, we define the mini-batch size and the number of training episodes as $|B|$ and $E_{\rm p}$, respectively. Thus, the complexity of the PDQN method can be denoted as $\mathcal{O}(|B| \times E_{\rm p} \times T \times (\sum\nolimits_{i=1}^{i=\mathscr{A}-1}\mathscr{A}_{i}\mathscr{A}_{i+1}+\sum\nolimits_{j=1}^{j=\mathscr{C}-1}\mathscr{C}_{j}\mathscr{C}_{j+1}))$~\cite{Guo2023}. In addition, let $\mathscr{P}_{\rm a}$ be the length of the parameters of the policy network and $\psi$ be the population size of the GA. Thus, the complexity of GA can be expressed as $\mathcal{O}(E_{\rm p} \times \psi \times \mathscr{P}_{\rm a})$. Consequently, the computation complexity of the IPDQN method is $\mathcal{O}((|B| \times E_{\rm p} \times T \times (\sum\nolimits_{i=1}^{i=\mathscr{A}-1}\mathscr{A}_{i}\mathscr{A}_{i+1}+\sum\nolimits_{j=1}^{j=\mathscr{C}-1}\mathscr{C}_{j}\mathscr{C}_{j+1})) + E_{\rm p} \times \psi \times \mathscr{P}_{\rm a})$

\par Based on the complexity analysis above, the complexity of the proposed AO-IPDQN approach can be denoted as $\mathcal{O}(Iter_{max} \times M_{\rm r} \times M_{\rm c} \times 2^{b}+(|B| \times E_{\rm p} \times T \times (\sum\nolimits_{i=1}^{i=\mathscr{A}-1}\mathscr{A}_{i}\mathscr{A}_{i+1}+\sum\nolimits_{j=1}^{j=\mathscr{C}-1}\mathscr{C}_{j}\mathscr{C}_{j+1})) + E_{\rm p} \times \psi \times \mathscr{P}_{\rm a})$.


%
%
\vspace{+0.5 mm}
\section{Simulation Results}
\label{sec:Simulation Results And Analysis}
\par We evaluate the performance of the AO-IPDQN approach through simulations in this section. The simulation setup is described first, followed by a presentation and analysis of the obtained simulation results.

\subsection{Simulation Setup}
\par The simulation settings are detailed in this part, including the configuration of the simulation environment and the design of the network used in the proposed approach.


\subsubsection{Environment Details} 
\par We consider a target area of size 400 m $\times$ 400 m, within which a UAV is deployed to provide data collection and power transfer services to 10 ground IoTDs that are randomly distributed across the target area. Moreover, each row and column of the RIS contains 8 reflecting elements. Other major system parameters applied in the simulation are listed in Table~\ref{Simulation Parameters}.

\subsubsection{Network Design} 
\par In our proposed AO-IPDQN approach, both the policy and value networks consist of two hidden layers, in which the dimensions of the hidden layers are $[400,200]$. Moreover, the dimensions of the input and output layers in the policy network are $[\mathcal{S}.dim]$ and $[\mathcal{D}.dim\times\mathcal{C}_{d}.dim]$, where $\mathcal{S}.dim$, $\mathcal{D}.dim$, and $\mathcal{C}_{d}.dim$ represent the dimensions of the state space, discrete action space, and continuous action space, respectively. Likewise, we denote the dimensions of the input and output layers in the value network as $[s.dim+\mathcal{D}.dim \times \mathcal{C}_{d}.dim]$ and $[\mathcal{D}.dim]$, respectively. Note that these networks adopt a fully connected layer with ReLU and Tanh activation functions. Moreover, we set the mini-batch size to 128 and the size of the replay memory to $10^6$. Moreover, the population size $\psi$ and the mutation rate $mu_{p}$ of the AO-IPDQN approach are 10 and 0.9.

\begin{table}[t]
\caption{Simulation Parameters.}
\label{Simulation Parameters}
\centering
\setlength{\tabcolsep}{6.3mm}{
\begin{tabular}{@{}@{\extracolsep{\fill}}llll@{}}
\toprule
Notation         & Value         & Notation         & Value         \\ \midrule
$X_{\rm max}$        & 400 m                & $Y_{\rm max}$        & 400 m                \\
$x_{\rm max}$        & 20 m                 & $y_{\rm max}$        & 20 m                 \\
$M_{\rm r}$          & 8                   & $M_{\rm c}$          & 8                   \\
$t$              & 1 s                  & $T$              & 120 s                \\
$P_{\rm U}$          & 30 dbm               & $P_{n}^{\rm D}$      & -30 dBm             \\
$d_{\rm r}$           & $\frac{\lambda}{2}$ & 
$d_{\rm c}$          & $\frac{\lambda}{2}$ \\ 
$r_{n}^{\rm RD}$     & 1                   & $r_{n}^{\rm UD}$     & 1                   \\
$\alpha_{0}$          & 0.001                 & $\sigma^{2}$     & -100 dBm            \\
$P_{\rm s}$          & 79.85               & $P_{\rm m}$          & 88.63               \\
$U_{\rm r}$          & 120 m/s             &$V_{\rm h}$           & 4.03                \\
$d_{0}$          & 0.6                 & $\rho_{\rm a}$       & 1.225 kg/$\rm{m^{3}}$     \\
$z$                & 0.05              & $G$              & 0.503 $\rm{m^{2}}$       \\ \bottomrule
\end{tabular}}
\end{table}

\begin{figure}[t]
    \centering
    \includegraphics[width=\linewidth]{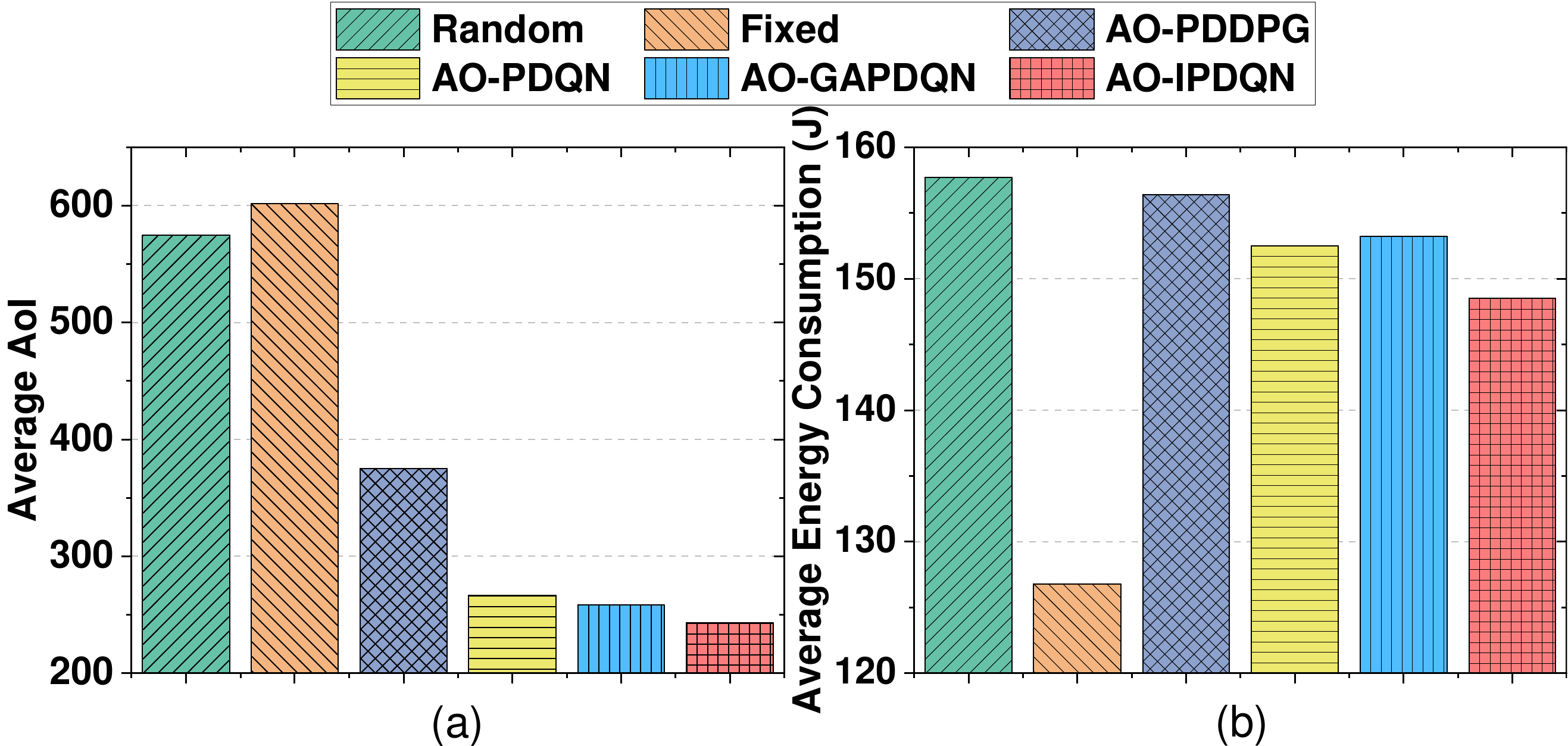}
    \caption{The optimization objective values obtained by different methods: (a) Average AoI and (b) Average UAV energy consumption.}
    \label{fig: Optimization Values}
\end{figure}

\begin{figure}[t]
\centering
\includegraphics[width=\linewidth]{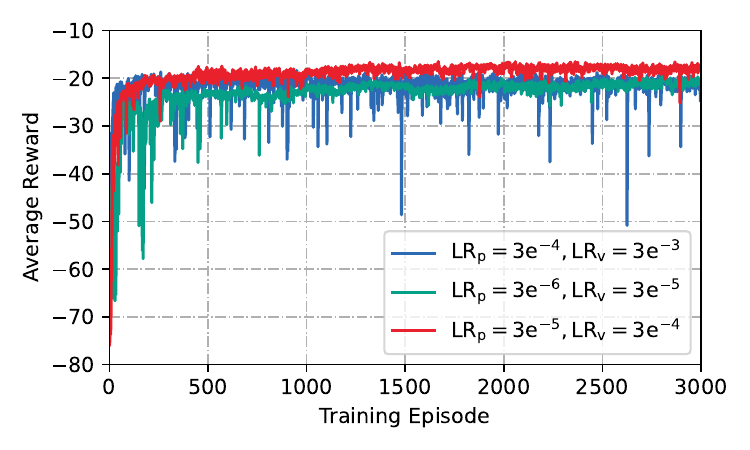}
\caption{The convergence curves of AO-IPDQN with different learning rates.}
\label{fig:ConvergenceLR}
\end{figure}

\begin{figure*}[htbp]
\centering
\subfigure[]{
\includegraphics[width=0.31\linewidth]{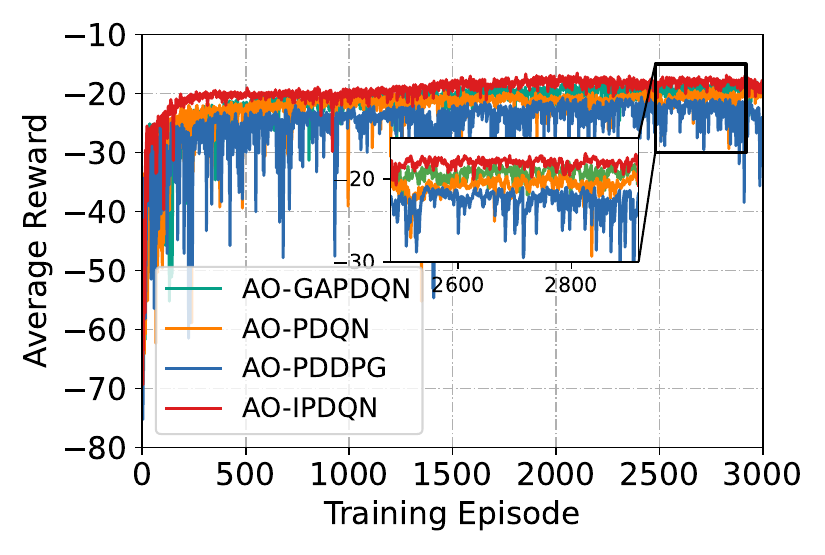}
}\label{subfig: convergence_reward}
\subfigure[]{
\includegraphics[width=0.31\linewidth]{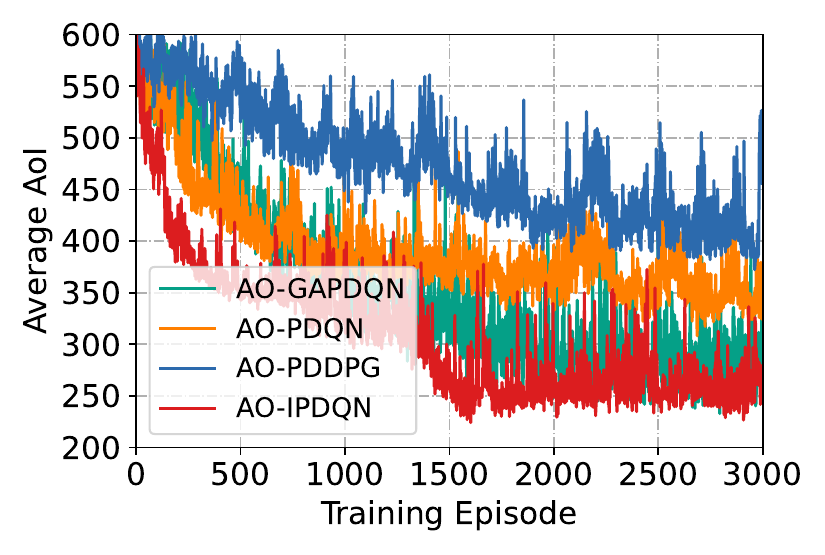}
}\label{subfig: Convergence AoI}
\subfigure[]{
\includegraphics[width=0.31\linewidth]{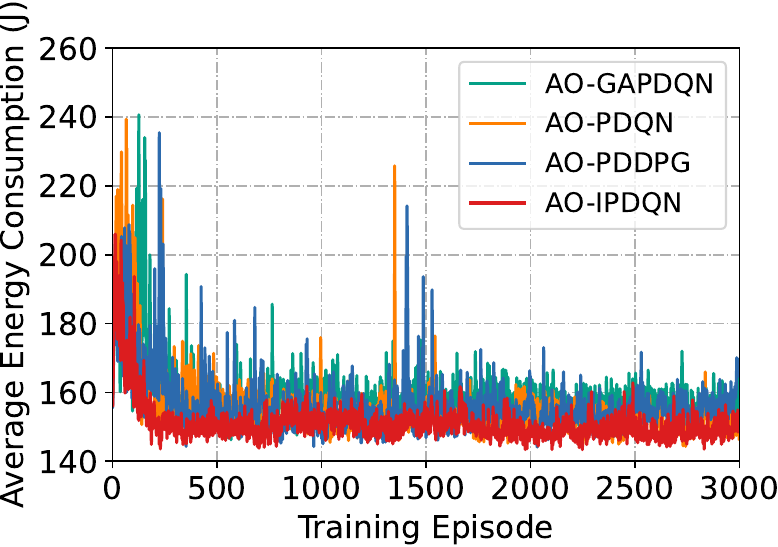}
}\label{subfig: Convergence Energy}
\caption{Convergence curves of different algorithms: (a) Average Reward, (b) Average AoI, and (c) Average UAV energy consumption.}
\label{fig: Convergence performance}
\end{figure*}

\subsection{Benchmark Methods}
\label{Benchmark Schemes}
\par To evaluate the effectiveness of our proposed AO-IPDQN approach, the several following comparison methods are considered.
\begin{itemize}
    \item \textit{Random Method}: In this method, the UAV fly randomly within the target area. Moreover, the RIS phase shifts, charging time allocation, and IoTD scheduling are set randomly while satisfying their constraints.
    \item \textit{Fixed Method}: In this method, the UAV flies along the diagonal of the target area at maximum endurance speed, and the UAV charging time is set to $0.5$ s. Moreover, each IoTD uploads its data in a dedicated time slot following a predetermined order, and the phase shifts of all the RIS reflecting elements are configured to $\pi/2$.
\end{itemize}

\par In addition to the traditional comparison methods above, we compare the proposed AO-IPDQN approach with the following DRL-based approaches.
\begin{itemize}
    \item \textit{AO-RIS and GA-improved PDQN Approach (AO-GAPDQN Approach)}: This approach optimizes the RIS phase shifts using the AO-RIS method. Moreover, a GA-improved PDQN method without the PER mechanism is adopted to optimize the UAV trajectory, charging time allocation, and IoTD scheduling~\cite{Khadka2018}.
    \item \textit{AO-RIS and PDQN Approach (AO-PDQN Approach)}: This approach employs the AO-RIS method to optimize the RIS phase shifts. Moreover, the conventional PDQN method is used to optimize the UAV trajectory, charging time allocation, and IoTD scheduling~\cite{Xiong2018}.
    \item \textit{AO-RIS and Parameterized Deep Deterministic Policy Gradient Approach (AO-PDDPG Approach)}: This approach configures the RIS phase shifts by the AO-RIS method. Moreover, the parameterized deep deterministic policy gradient (PDDPG) method~\cite{Hausknecht2016} is adopted to optimize the UAV trajectory, charging time allocation, and IoTD scheduling. Notably, although both the PDQN and PDDPG methods are designed to handle the hybrid action space, the PDDPG method differs by directly outputting all the discrete and continuous actions through a policy network.
\end{itemize}

\subsection{Optimization Results}
\label{Optimization Results}

\par In this part, we provide a detailed analysis of the proposed Ao-IPDQN approach in terms of the optimization performance and convergence performance.

\subsubsection{Algorithm Performance Evaluation}
\par Fig.~\ref{fig: Optimization Values} shows the optimization objective values obtained by different methods. As can be seen, the proposed AO-IPDQN approach achieves the best AoI performance and suboptimal UAV energy consumption. The phenomenon occurs because the policy network population maintained by the AO-IPDQN approach can more thoroughly explore the action space and discover better solutions due to its improved exploration capability. Moreover, the introduction of the PER mechanism contributes to faster learning while simultaneously improving learning quality, thereby improving the decision-making performance of the AO-IPDQN approach. In addition, we observe that the AO-PDDPG method performs the worst among the DRL-based approaches. This is because the policy network in the AO-PDDPG method generates both discrete and continuous actions simultaneously, which neglects the potential interdependencies between these action types and results in poor decision-making. Furthermore, the random method performs poorly as all optimization parameters are set randomly, and the fixed method achieves the lowest energy consumption because the UAV flies at maximum endurance speed. In summary, the AO-IPDQN approach achieves the highest AoI performance while simultaneously yielding a significant reduction in UAV energy consumption, which is crucial for the considered system based on an energy-constrained UAV platform.

\subsubsection{Convergence Performance Evaluation}
\par In this part, we analyze the impact of different learning rats on the proposed Ao-IPDQN approach, and then compare the convergence performance of all methods.

\par \textit{1) Impact of Learning Rate}: Fig.~\ref{fig:ConvergenceLR} shows the convergence performance of our proposed AO-IPDQN approach with different learning rates. We can observe that the AO-IPDQN approach can achieve better reward and faster convergence speed when the learning rates of the policy network and value network are set to $\rm LR_{p}=3\times 10^{-5}$ and $\rm LR_{v}=3\times 10^{-4}$, respectively. This phenomenon occurs because a small learning rate results in minimal parameter updates, which causes the agent to learn slowly and potentially get stuck in the local optimal.In contrast, when the learning rate is set too large, the parameter updates become too aggressive, which may make the agent overlook the optimal policy and lead to instability or failure to converge to the optimal solution. As such, the learning rate plays a critical role in balancing stability and exploration.

\par \textit{2) Comparison with Other Methods}: Fig.~\ref{fig: Convergence performance}(a) demonstrates the convergence performance of different methods. As can be seen, the reward curves of all methods oscillate significantly at the beginning. This phenomenon occurs because the agent lacks experience and fails to learn the constraints of UAV flight actions during the initial training phase, causing the UAV to frequently fly outside the target area and resulting in penalties for the agent. With increasing training episodes, the reward curves of all methods gradually converge, which indicates that all methods progressively learn a more stable strategy.Among the methods, our proposed AO-IPDQN approach achieves higher reward and converges more quickly than the other methods. This is because the policy network population helps maintain training stability while facilitating the discovery of better solutions. Moreover, the PER mechanism improves learning efficiency, thereby accelerating the convergence of the AO-IPDQN approach. In addition, Fig.~\ref{fig: Convergence performance}(b) and Fig.~\ref{fig: Convergence performance}(c) present the training curves of the average AoI and energy consumption. We can observe that although the AO-IPDQN and AO-GAPDQN approaches exhibit similar performance in terms of AoI metric, the proposed AO-IPDQN approach obtains lower UAV energy consumption, indicating that the AO-IPDQN approach is more effective in designing the UAV trajectory.

\begin{figure}
    \centering
    \includegraphics[width=\linewidth]{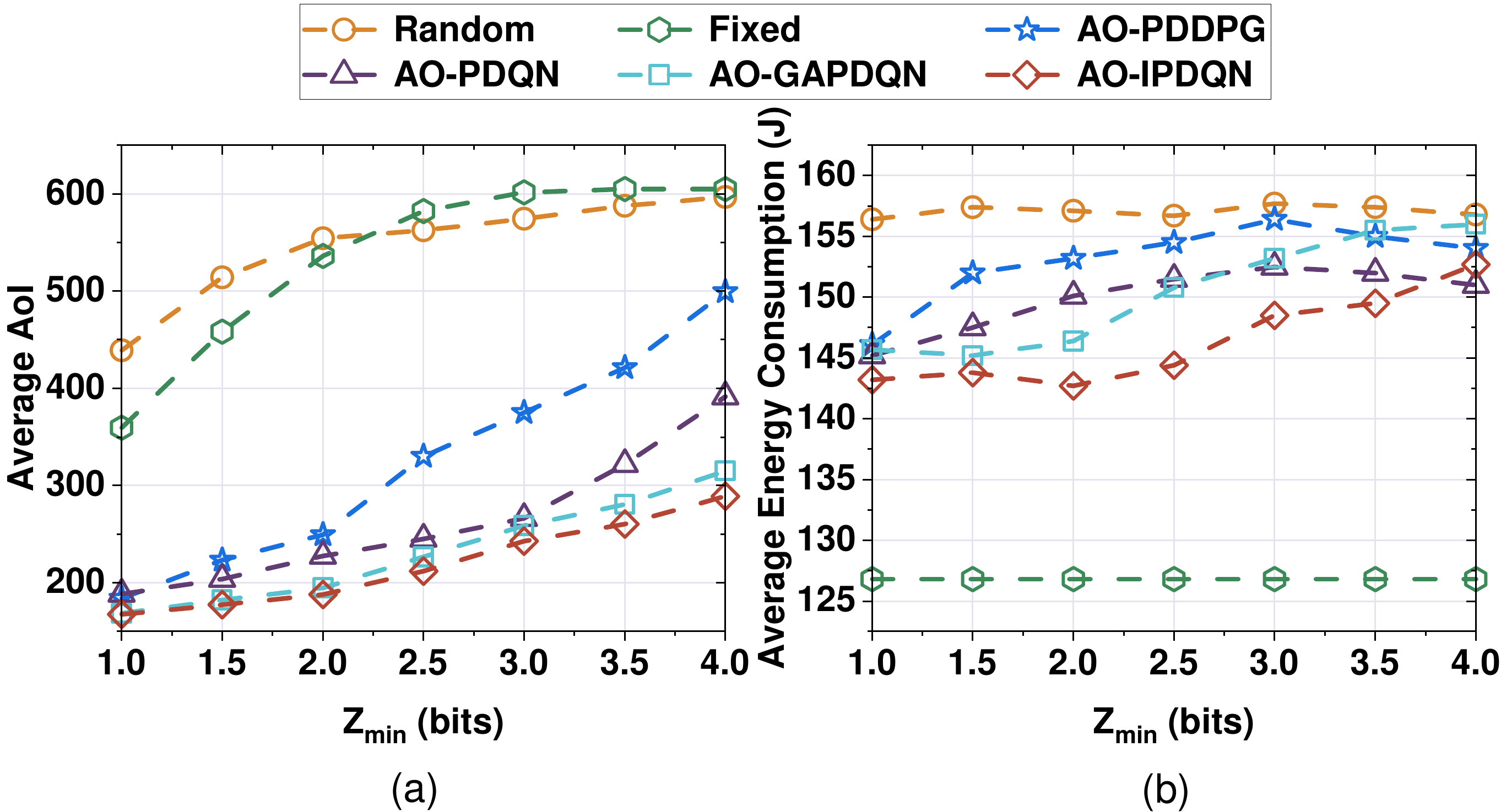}
    \caption{The impact of minimum data size threshold on different optimization objectives: (a) Average AoI and (b) Average UAV energy consumption.}
    \label{fig: Impact of Zmin}
\end{figure}

\subsection{Impact of System Settings}
\par In this part, we analyze the impact of the minimum data size threshold and energy buffer capacity on the optimization objectives.

\par \textit{1) Impact of Minimum Data Size Threshold $Z_{\rm min}$}: Fig.~\ref{fig: Impact of Zmin} shows the impact of the minimum data size threshold $Z_{min}$ on the AoI of IoTDs and the energy consumption of the UAV. As can be seen, the average AoI and average UAV energy consumption of the proposed AO-IPDQN approach gradually increases with the increasing minimum data size threshold. This phenomenon occurs because the conditions for uploading data by IoTDs become more stringent as $Z_{\rm min}$ increases, which leads to higher AoI of the system. Meanwhile, the UAV should fly towards the vicinity of the IoTDs for better channel conditions to improve the probability of collecting data successfully, which leads to higher UAV energy consumption. Moreover, we observe that our proposed AO-IPDQN approach outperforms most comparison methods with respect to average AoI and average UAV energy consumption with different $Z_{\rm min}$, which can be attributed to its effective exploration of the hybrid action space.

\begin{figure}
    \centering
    \includegraphics[width=\linewidth]{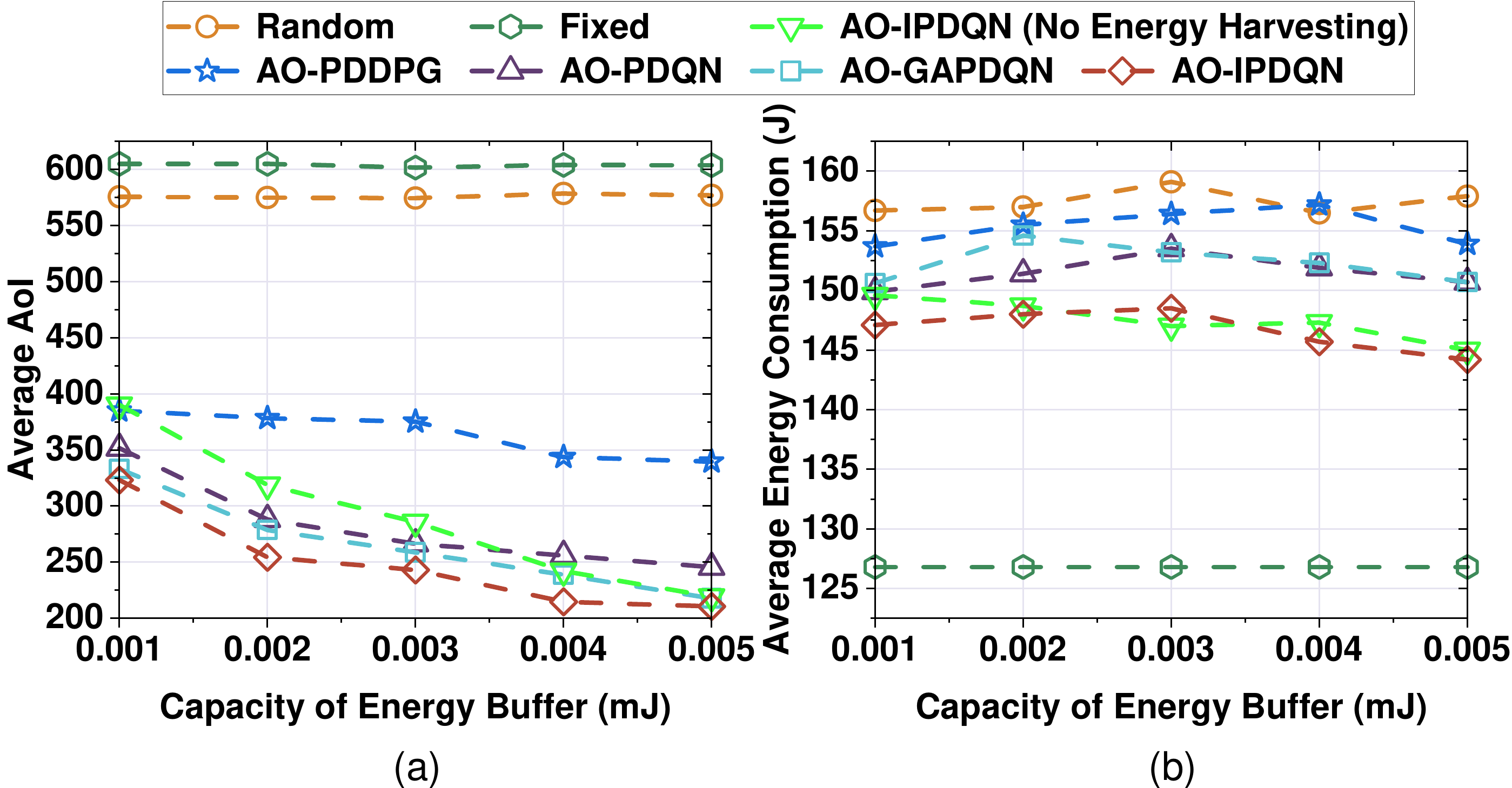}
    \caption{The impact of the energy buffer capacity of IoTD on different optimization objectives: (a) Average AoI and (b) Average UAV energy consumption.}
    \label{fig: Impact of Emax}
\end{figure}

\par \textit{2) Impact of Energy Buffer Capacity $E_{\rm max}$}: Fig.~\ref{fig: Impact of Emax} demonstrates the impact of the energy buffer capacity $E_{\rm max}$ on average AoI and UAV energy consumption. As can be seen, the AoI metric under most methods shows a decreasing trend with the growth of energy buffer capacity, primarily because a larger buffer capacity allows IoTDs to upload data more frequently and reliably. However, the AoI under the random method and the fixed method is almost unaffected by the changing energy buffer capacity. This phenomenon occurs because both methods lack adaptability in allocating resources and scheduling IoTDs, which decreases the data upload success rate and leaves substantial energy unused in the energy buffer of IoTDs. Moreover, the AoI performance of the AO-IPDQN approach without applying the energy harvesting technology is poor when the energy buffer capacity is small. This is due to the fact that IoTDs fail to upload data in later stages due to insufficient sustainable energy supply. In summary, compared to other methods, the AO-IPDQN approach demonstrates better AoI performance, particularly under the challenging condition of low energy buffer capacity. This indicates that the AO-IPDQN approach can effectively design the UAV power transfer strategy, ensuring that IoTDs can continuously receive energy supplementation under limited conditions, thereby enabling efficient data uploads.

\begin{figure}
    \centering
    \includegraphics[width=0.85\linewidth]{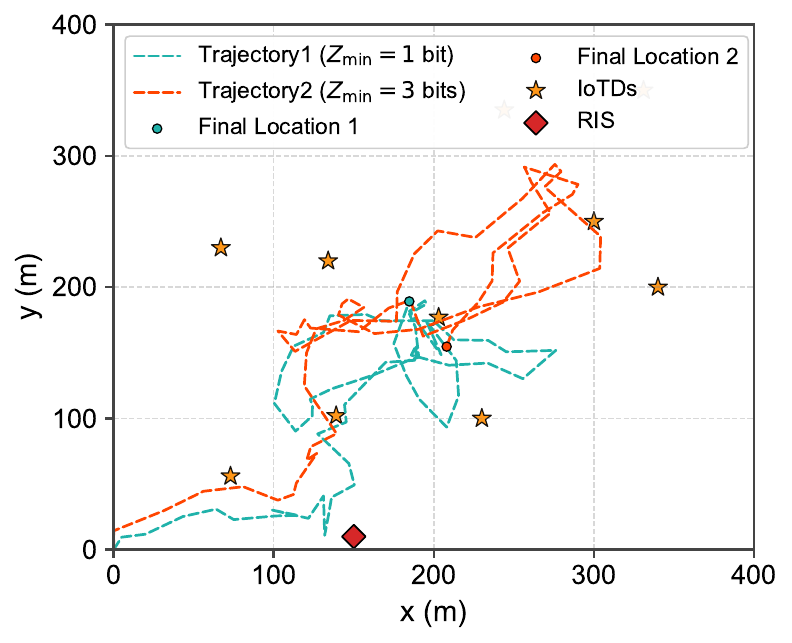}
    \caption{The trajectories of the UAV under different minimum data size thresholds when the UAV starts at the same location.}
    \label{fig: Trajectory}
\end{figure}

\subsection{UAV Trajectory Visualization}

\par Fig.~\ref{fig: Trajectory} shows the UAV trajectories under different minimum data size thresholds. As can be seen, when the threshold is relatively small, the UAV prefers to remain close to the RIS instead of extending its flight range. This phenomenon occurs because the UAV hovering near the RIS can already meet the minimum data size threshold and achieve satisfactory AoI performance. In contrast, when the threshold becomes large, the UAV expands its activity coverage to approach the IoTD-intensive areas for achieving better LoS link quality. This is because LoS links typically offer higher gains than RIS reflection links, which is beneficial to improve the communication rates between the UAV and IoTDs and consequently contributes to reducing the overall AoI of the system. In summary, the aforementioned results indicate that the AO-IPDQN approach can design reasonable UAV trajectory schemes according to different system requirements.


%
%
\section{Conclusion}
\label{sec:Conclusion}
\par In this paper, we have studied an RIS-assisted UAV-enabled data collection and wireless power transfer system. Moreover, we have formulated a multi-objective optimization problem which aims to minimize the AoI and UAV energy consumption by jointly optimizing the RIS phase shifts, UAV trajectory, charging time allocation, and IoTD scheduling. To tackle this mixed-integer non-convex problem, we have proposed an AO-IPDQN approach, which combines the AO method to reduce the dimension of the action space and utilizes the IPDQN method to deal with the hybrid action space in a more efficient way. Simulation results have revealed that the AO-IPDQN approach achieves substantial reductions in AoI and UAV energy consumption compared to other comparison methods. In particular, the proposed AO-IPDQN approach has demonstrated satisfactory performance under high data size thresholds and low-capacity energy buffers. In addition, the AO-IPDQN approach can design different UAV trajectories based on environmental changes, demonstrating its superior adaptability and robustness.

\bibliographystyle{IEEEtran}
\normalem
\bibliography{cite_paper2}

\vspace{-10mm}

\begin{IEEEbiography}[{\includegraphics[width=1in,height=1.25in,clip,keepaspectratio]{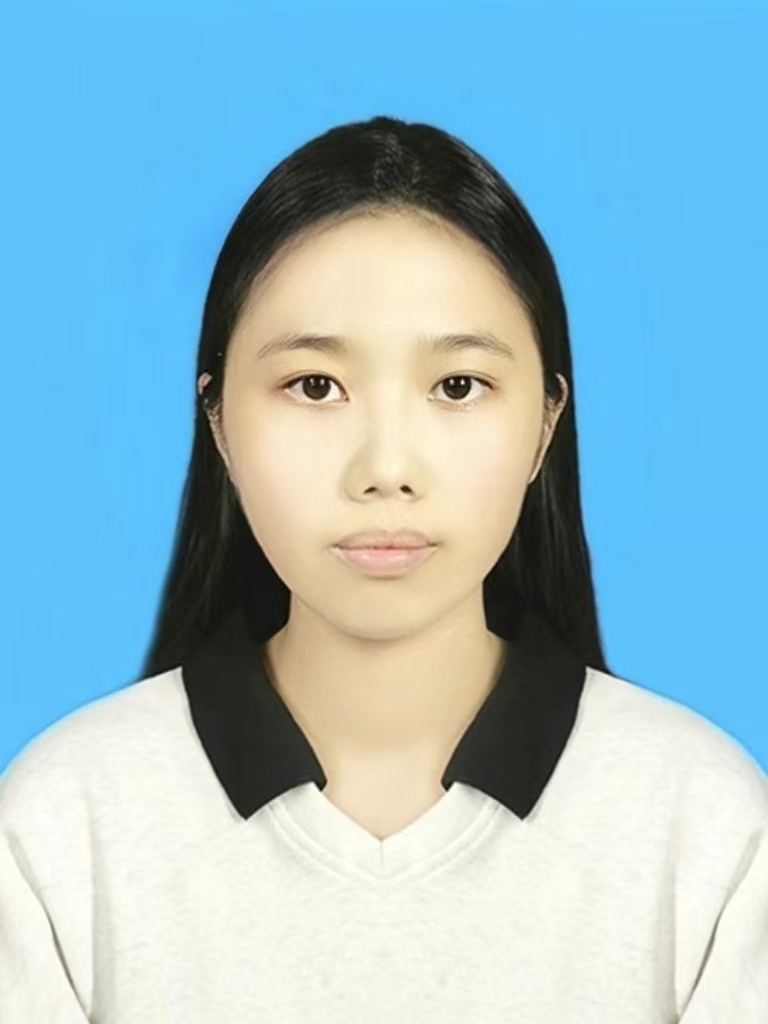}}]{Wenwen Xie} received the B.S. degree in Computer Science and Technology from Hefei University of Technology, Hefei, China, in 2022. She received the M.S. degree in Computer Science and Technology at Jilin University, Changchun, China, in 2025. She is currently working toward the Ph.D. degree in Computer Science and Technology at Jilin University, Changchun, China. Her research interests include UAV communications, RIS beamforming, ISAC, and deep reinforcement learning.

\end{IEEEbiography}

\vspace{-10mm}

\begin{IEEEbiography}[{\includegraphics[width=1in,height=1.25in,clip,keepaspectratio]{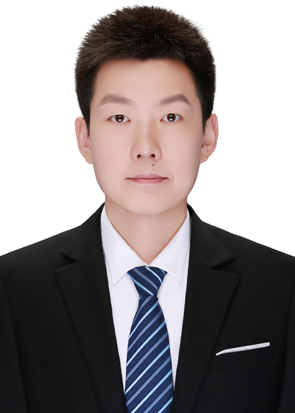}}]{Geng Sun} (Senior Member, IEEE) received the B.S. degree in communication engineering from Dalian Polytechnic University, and the Ph.D. degree in computer science and technology from Jilin University, in 2011 and 2018, respectively. He was a Visiting Researcher with the School of Electrical and Computer Engineering, Georgia Institute of Technology, USA. He is a Professor in the College of Computer Science and Technology at Jilin University. Currently, he is working as a visiting scholar at the College of Computing and Data Science, Nanyang Technological University, Singapore. He has published over 100 high-quality papers, including IEEE TMC, IEEE JSAC, IEEE/ACM ToN, IEEE TWC, IEEE TCOM, IEEE TAP, IEEE IoT-J, IEEE TIM, IEEE INFOCOM, IEEE GLOBECOM, and IEEE ICC. He serves as the Associate Editors of IEEE Communications Surveys \& Tutorials, IEEE Transactions on Vehicular Technology, IEEE Transactions on Network Science and Engineering, and IEEE Networking Letters. He serves as the Lead Guest Editor of Special Issues for IEEE Transactions on Network Science and Engineering, IEEE Internet of Things Journal, IEEE Networking Letters. He also serves as the Guest Editor of Special Issues for IEEE Transactions on Services Computing, IEEE Communications Magazine, and IEEE Open Journal of the Communications Society. His research interests include UAV communications and networking, mobile edge computing (MEC), intelligent reflecting surface (IRS), generative AI, and deep reinforcement learning.
\end{IEEEbiography}

\vspace{-10mm}
\begin{IEEEbiography}
[{\includegraphics[width=1in,height=1.25in,clip,keepaspectratio]{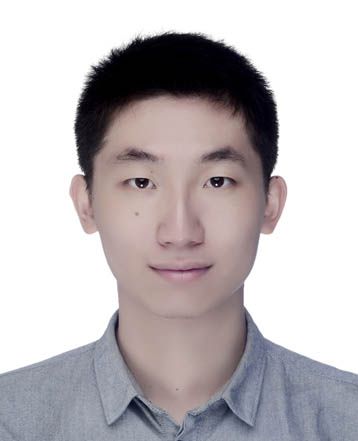}}]
{Jiahui Li} (Member, IEEE) received his B.S. in Software Engineering, and M.S. and Ph.D. in Computer Science and Technology from Jilin University, Changchun, China, in 2018, 2021, and 2024, respectively. He was a visiting Ph.D. student at the Singapore University of Technology and Design (SUTD). He currently serves as an assistant researcher in the College of Computer Science and Technology at Jilin University. His current research focuses on integrated air-ground networks, UAV networks, wireless energy transfer, and optimization.
\end{IEEEbiography}

\vspace{-10mm}

\begin{IEEEbiography}[{\includegraphics[width=1in,height=1.25in,clip,keepaspectratio]{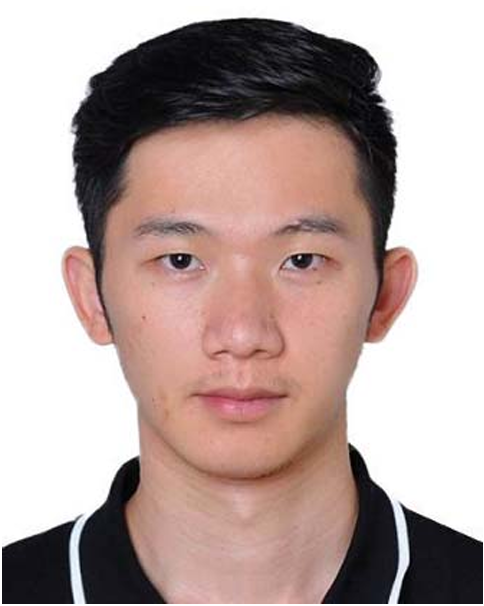}}]{Jiacheng Wang} received the Ph.D. degree from the School of Communication and Information Engineering, Chongqing University of Posts and Telecommunications, Chongqing, China. He is currently a Research Associate in computer science and engineering with Nanyang Technological University, Singapore. His research interests include wireless sensing, semantic communications, and metaverse.
\end{IEEEbiography}
\vspace{-10mm}
\begin{IEEEbiography}
[{\includegraphics[width=1in,height=1.25in,clip,keepaspectratio]{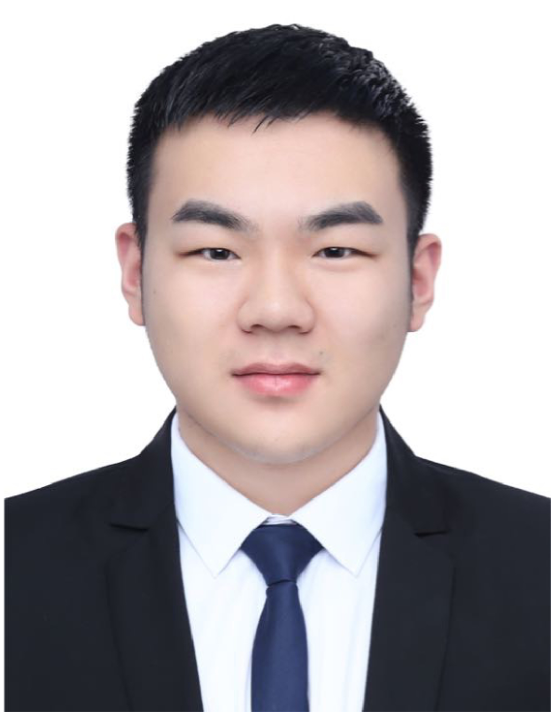}}]{Yinqiu Liu} received B.Eng. degree from Nanjing University of Posts and Telecommunications, China in 2020 and M.Sc degree from the University of California, Los Angeles in 2022. He is currently a Ph.D. candidate at the College of Computing and Data Science, Nanyang Technological University, Singapore. His current research interests include wireless communications, mobile AIGC, and generative AI. E-mail: yinqiu001@e.ntu.edu.sg.
\end{IEEEbiography}
\vspace{-10mm}
\begin{IEEEbiography}[{\includegraphics[width=1in,height=1.25in,clip,keepaspectratio]{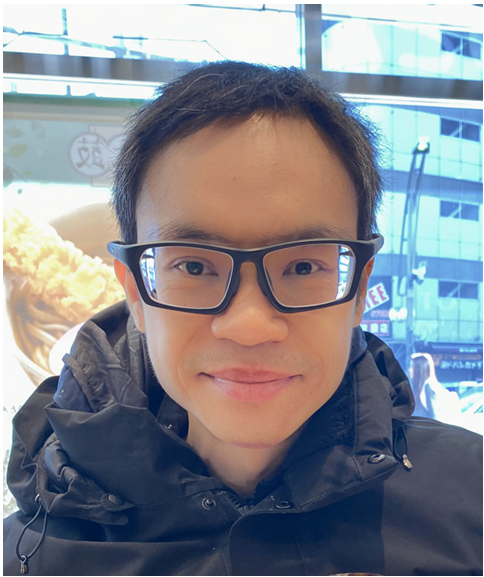}}]{Dusit Niyato} (Fellow, IEEE) received the B.Eng. degree from the King Mongkuts Institute of Technology Ladkrabang (KMITL), Thailand, in 1999, and the Ph.D. degree in electrical and computer engineering from the University of Manitoba, Canada, in 2008. He is currently a Professor with the School of Computer Science and Engineering, Nanyang Technological University, Singapore. His research interests include the Internet of Things (IoT), machine learning, and incentive mechanism design. 
\end{IEEEbiography}

\vspace{-10mm}
\begin{IEEEbiography}[{\includegraphics[width=1in,height=1.25in,clip,keepaspectratio]{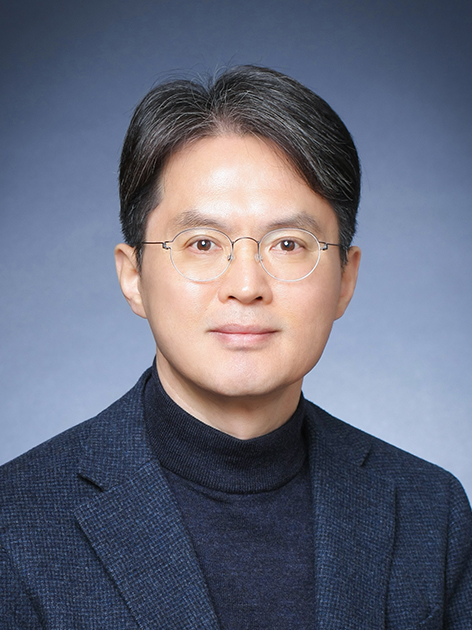}}]{Dong In Kim} (Fellow, IEEE) received the Ph.D. degree in electrical engineering from the University of Southern California, Los Angeles, CA, USA, in 1990. He was a Tenured Professor with the School of Engineering Science, Simon Fraser University, Burnaby, BC, Canada. He is currently a Distinguished Professor with the College of Information and Communication Engineering, Sungkyunkwan University, Suwon, South Korea. He is a Fellow of the Korean Academy of Science and Technology and a Member of the National Academy of Engineering of Korea. He was the first recipient of the NRF of Korea Engineering Research Center in Wireless Communications for RF Energy Harvesting from 2014 to 2021. He received several research awards, including the 2023 IEEE ComSoc Best Survey Paper Award and the 2022 IEEE Best Land Transportation Paper Award. He was selected the 2019 recipient of the IEEE ComSoc Joseph LoCicero Award for Exemplary Service to Publications. He was the General Chair of the IEEE ICC 2022, Seoul. Since 2001, he has been serving as an Editor, an Editor at Large, and an Area Editor of Wireless Communications I for IEEE Transactions on Communications. From 2002 to 2011, he served as an Editor and a Founding Area Editor of Cross-Layer Design and Optimization for IEEE Transactions on Wireless Communications. From 2008 to 2011, he served as the Co-Editor- in-Chief for the IEEE/KICS Journal of Communications and Networks. He served as the Founding Editor-in-Chief for the IEEE Wireless Communications Letters from 2012 to 2015. He has been listed as a 2020/2022 Highly Cited Researcher by Clarivate Analytics.
\end{IEEEbiography}

\begin{IEEEbiography}[{\includegraphics[width=1in,height=1.25in,clip,keepaspectratio]{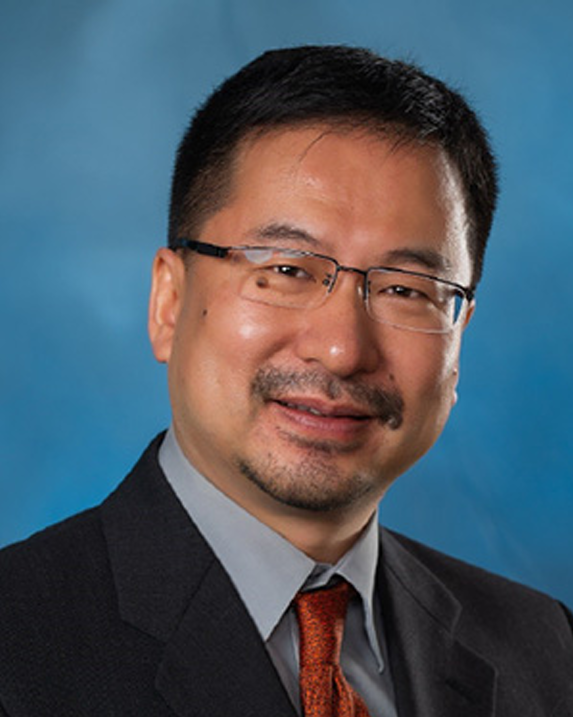}}]{Shiwen Mao} (Fellow, IEEE) is a Professor and the Earle C. Williams Eminent Scholar Chair, and the Director of the Wireless Engineering Research and Education Center, Auburn University, Auburn, AL, USA. His research interest includes wireless networks, multimedia communications, and smart grid. He received the IEEE ComSoc MMTC Outstanding Researcher Award in 2023, the IEEE ComSoc TC-CSR Distinguished Technical Achievement Award in 2019, and the NSF CAREER Award in 2010. He is a co-recipient of the 2022 Best Journal Paper Award of IEEE ComSoc eHealth Technical Committee, the 2021 Best Paper Award of Elsevier/Digital Communications and Networks (KeAi), the 2021 IEEE Internet of Things Journal Best Paper Award, the 2021 IEEE Communications Society Outstanding Paper Award, the IEEE Vehicular Technology Society 2020 Jack Neubauer Memorial Award, the 2018 ComSoc MMTCBestJournal Paper Award and the 2017 Best Conference Paper Award, the 2004 IEEE Communications Society Leonard G. Abraham Prize in the Field of Communications Systems, and several ComSoc technical committee and conference best paper/demo awards. He is the Editor-in-Chief of IEEE TRANSACTIONS ON COGNITIVE COMMUNICATIONS AND NETWORKING. He is a Distinguished Lecturer of IEEE Communications Society and the IEEE Council of RFID.

\end{IEEEbiography}

\end{document}